\documentclass[twocolumn,pre,aps,nofootinbib,superscriptaddress,showpacs,preprintnumbers]{revtex4-1}
\usepackage{graphicx,times}
\usepackage{bm,url}
\begin{document}
\preprint{NORDITA-2011-40}


\newcommand{\bec}[1]{\mbox{\boldmath $ #1$}}
\newcommand{\EQ}{\begin{equation}}
\newcommand{\EN}{\end{equation}}
\newcommand{\EQA}{\begin{eqnarray}}
\newcommand{\ENA}{\end{eqnarray}}
\newcommand{\eq}[1]{(\ref{#1})}
\newcommand{\EEq}[1]{Equation~(\ref{#1})}
\newcommand{\Eq}[1]{Eq.~(\ref{#1})}
\newcommand{\Eqs}[2]{Eqs.~(\ref{#1}) and~(\ref{#2})}
\newcommand{\Eqss}[2]{Eqs.~(\ref{#1})--(\ref{#2})}
\newcommand{\eqs}[2]{(\ref{#1}) and~(\ref{#2})}
\newcommand{\App}[1]{Appendix~\ref{#1}}
\newcommand{\Sec}[1]{Sect.~\ref{#1}}
\newcommand{\Secs}[2]{Sects.~\ref{#1} and \ref{#2}}
\newcommand{\Fig}[1]{Fig.~\ref{#1}}
\newcommand{\FFig}[1]{Figure~\ref{#1}}
\newcommand{\Figs}[2]{Figs.~\ref{#1} and \ref{#2}}
\newcommand{\Figss}[2]{Figs.~\ref{#1}--\ref{#2}}
\newcommand{\Tab}[1]{Table~\ref{#1}}
\newcommand{\Tabs}[2]{Tables~\ref{#1} and \ref{#2}}
\newcommand{\bra}[1]{\langle #1\rangle}
\newcommand{\bbra}[1]{\left\langle #1\right\rangle}
\newcommand{\mean}[1]{\overline #1}
\newcommand{\meanv}[1]{\bm{\overline #1}}
\newcommand{\pd}{\partial}
\newcommand{\meanrho}{\overline{\rho}}
\newcommand{\meanPhi}{\overline{\Phi}}
\newcommand{\tildeFFFF}{\tilde{\mbox{\boldmath ${\cal F}$}}{}}{}
\newcommand{\hatFFFF}{\hat{\mbox{\boldmath ${\cal F}$}}{}}{}
\newcommand{\meanFFFF}{\overline{\mbox{\boldmath ${\cal F}$}}{}}{}
\newcommand{\meanemf}{\overline{\cal E} {}}
\newcommand{\meanAAAA}{\overline{\mbox{\boldmath ${\mathsf A}$}} {}}
\newcommand{\meanSSSS}{\overline{\mbox{\boldmath ${\mathsf S}$}} {}}
\newcommand{\meanAAA}{\overline{\mathsf{A}}}
\newcommand{\meanSSS}{\overline{\mathsf{S}}}
\newcommand{\meanuu}{\overline{\mbox{\boldmath $u$}}{}}{}
\newcommand{\meanoo}{\overline{\mbox{\boldmath $\omega$}}{}}{}
\newcommand{\meanEMF}{\overline{\mbox{\boldmath ${\cal E}$}}{}}{}
\newcommand{\meanuxB}{\overline{\mbox{\boldmath $\delta u\times \delta B$}}{}}{}
\newcommand{\meanJB}{\overline{\mbox{\boldmath $J\cdot B$}}{}}{}
\newcommand{\meanAB}{\overline{\mbox{\boldmath $A\cdot B$}}{}}{}
\newcommand{\meanjb}{\overline{\mbox{\boldmath $j\cdot b$}}{}}{}
\newcommand{\meanAA}{\overline{\mbox{\boldmath $A$}}{}}{}
\newcommand{\meanBB}{\overline{\mbox{\boldmath $B$}}{}}{}
\newcommand{\meanEE}{\overline{\mbox{\boldmath $E$}}{}}{}
\newcommand{\meanFF}{\overline{\mbox{\boldmath $F$}}{}}{}
\newcommand{\meanFFf}{\overline{\mbox{\boldmath $F$}}_{\rm f}{}}{}
\newcommand{\meanFFm}{\overline{\mbox{\boldmath $F$}}_{\rm m}{}}{}
\newcommand{\hatFF}{\hat{\mbox{\boldmath $F$}}{}}{}
\newcommand{\meanGG}{\overline{\mbox{\boldmath $G$}}{}}{}
\newcommand{\tildeGG}{\tilde{\mbox{\boldmath $G$}}{}}{}
\newcommand{\meanJJ}{\overline{\mbox{\boldmath $J$}}{}}{}
\newcommand{\meanUU}{\overline{\bm{U}}}
\newcommand{\meanVV}{\overline{\bm{V}}}
\newcommand{\meanWW}{\overline{\mbox{\boldmath $W$}}{}}{}
\newcommand{\hatWW}{\hat{\mbox{\boldmath $W$}}{}}{}
\newcommand{\meanQQ}{\overline{\mbox{\boldmath $Q$}}{}}{}
\newcommand{\meanA}{\overline{A}}
\newcommand{\meanB}{\overline{B}}
\newcommand{\meanb}{\tilde{b}}
\newcommand{\meanj}{\tilde{j}}
\newcommand{\meanC}{\overline{C}}
\newcommand{\meanF}{\overline{F}}
\newcommand{\meanFm}{\overline{F}_{\rm m}}
\newcommand{\meanFf}{\overline{F}_{\rm f}}
\newcommand{\meanh}{\overline{h}}
\newcommand{\meanhm}{\overline{h}_{\rm m}}
\newcommand{\meanhf}{\overline{h}_{\rm f}}
\newcommand{\meanG}{\overline{G}}
\newcommand{\meanH}{\overline{H}}
\newcommand{\meanU}{\overline{U}}
\newcommand{\meanR}{\overline{\rho}}
\newcommand{\meanJ}{\overline{J}}
\newcommand{\means}{\overline{s}}
\newcommand{\meanp}{\overline{p}}
\newcommand{\meanS}{\overline{S}}
\newcommand{\meanT}{\overline{T}}
\newcommand{\meanW}{\overline{W}}
\newcommand{\meanQ}{\overline{Q}}
\newcommand{\meanFFF}{\overline{\cal F}}
\newcommand{\hatAA}{\hat{\bm{A}}}
\newcommand{\hatBB}{\hat{\bm{B}}}
\newcommand{\hatJJ}{\hat{\mbox{\boldmath $J$}}{}}{}
\newcommand{\hatOO}{\hat{\bm{\Omega}}}
\newcommand{\hatEMF}{\hat{\mbox{\boldmath ${\cal E}$}}{}}{}
\newcommand{\emf}{{\cal E}}{}
\newcommand{\hatA}{\hat{A}}
\newcommand{\hatB}{\hat{B}}
\newcommand{\hatJ}{\hat{J}}
\newcommand{\hatU}{\hat{U}}
\newcommand{\alphaK}{\alpha_{\it K}}
\newcommand{\alphaM}{\alpha_{\it M}}
\newcommand{\kef}{k_{\rm f}}
%
%
\newcommand{\teps}{\tilde{\epsilon} {}}
\newcommand{\Oh}{\hat{\Omega}}
\newcommand{\zh}{\hat{z}}
\newcommand{\PC}{{\sc Pencil Code}~}
%
%
\newcommand{\pphi}{\hat{\bm{\phi}}}
\newcommand{\ppom}{\bm{\hat{\varpi}}}
\newcommand{\eee}{\hat{\mbox{\boldmath $e$}} {}}
\newcommand{\nnn}{\hat{\mbox{\boldmath $n$}} {}}
\newcommand{\rrr}{\hat{\mbox{\boldmath $r$}} {}}
\newcommand{\vvv}{\hat{\mbox{\boldmath $v$}} {}}
\newcommand{\xxx}{\hat{\mbox{\boldmath $x$}} {}}
\newcommand{\yyy}{\hat{\mbox{\boldmath $y$}} {}}
\newcommand{\zzz}{\hat{\mbox{\boldmath $z$}} {}}
\newcommand{\ttt}{\hat{\mbox{\boldmath $\theta$}} {}}
\newcommand{\OOO}{\hat{\mbox{\boldmath $\Omega$}} {}}
\newcommand{\ooo}{\hat{\mbox{\boldmath $\omega$}} {}}
\newcommand{\BBBB}{\hat{\mbox{\boldmath $B$}} {}}
\newcommand{\meanBBhat}{\hat{\overline{\bm{B}}}}
\newcommand{\meanJJhat}{\hat{\overline{\bm{J}}}}
\newcommand{\meanBhat}{\hat{\overline{B}}}
\newcommand{\meanJhat}{\hat{\overline{J}}}
\newcommand{\Bhat}{\hat{B}}
\newcommand{\Bh}{\hat{B}}
%
%
\newcommand{\nullvector}{{\bf0}}
\newcommand{\gggg}{\mbox{\boldmath $g$} {}}
\newcommand{\ddd}{\mbox{\boldmath $d$} {}}
\newcommand{\rr}{\mbox{\boldmath $r$} {}}
\newcommand{\yy}{\mbox{\boldmath $y$} {}}
\newcommand{\zz}{\mbox{\boldmath $z$} {}}
\newcommand{\vv}{\mbox{\boldmath $v$} {}}
\newcommand{\ww}{\mbox{\boldmath $w$} {}}
\newcommand{\mm}{\mbox{\boldmath $m$} {}}
\newcommand{\PP}{\mbox{\boldmath $P$} {}}
\newcommand{\bp}{\mbox{\boldmath $p$} {}}
\newcommand{\II}{\mbox{\boldmath $I$} {}}
\newcommand{\RR}{\mbox{\boldmath $R$} {}}
\newcommand{\kk}{\bm{k}}
\newcommand{\pp}{\bm{p}}
\newcommand{\qq}{\bm{q}}
\newcommand{\xx}{\bm{x}}
\newcommand{\KK}{\mbox{\boldmath $K$} {}}
\newcommand{\uu}{\mbox{\boldmath $u$} {}}
\newcommand{\UU}{\mbox{\boldmath $U$} {}}
\newcommand{\sss}{\mbox{\boldmath $s$} {}}
\newcommand{\bb}{\mbox{\boldmath $b$} {}}
\newcommand{\BB}{\mbox{\boldmath $B$} {}}
\newcommand{\EE}{\mbox{\boldmath $E$} {}}
\newcommand{\jj}{\mbox{\boldmath $j$} {}}
\newcommand{\JJ}{\mbox{\boldmath $J$} {}}
\newcommand{\SSS}{\mbox{\boldmath $S$} {}}
\newcommand{\AAA}{\mbox{\boldmath $A$} {}}
\newcommand{\aaaa}{\mbox{\boldmath $a$} {}}
\newcommand{\ee}{\mbox{\boldmath $e$} {}}
\newcommand{\nn}{\mbox{\boldmath $n$} {}}
\newcommand{\ff}{\mbox{\boldmath $f$} {}}
\newcommand{\hh}{\mbox{\boldmath $h$} {}}
\newcommand{\FF}{\mbox{\boldmath $F$} {}}
\newcommand{\EEE}{\mbox{\boldmath ${\cal E}$} {}}
\newcommand{\FFF}{\mbox{\boldmath ${\cal F}$} {}}
\newcommand{\TT}{{\bm{T}}}
\newcommand{\MM}{\mbox{\boldmath $M$} {}}
\newcommand{\GG}{\mbox{\boldmath $G$} {}}
\newcommand{\WW}{\mbox{\boldmath $W$} {}}
\newcommand{\QQ}{\mbox{\boldmath $Q$} {}}
\newcommand{\grav}{\mbox{\boldmath $g$} {}}
\newcommand{\nab}{\mbox{\boldmath $\nabla$} {}}
\newcommand{\OO}{\bm{\Omega}}
\newcommand{\oo}{\mbox{\boldmath $\omega$} {}}
\newcommand{\pom}{\mbox{\boldmath $\varpi$} {}}
\newcommand{\ttau}{\mbox{\boldmath $\tau$} {}}
\newcommand{\LL}{\mbox{\boldmath $\Lambda$} {}}
\newcommand{\mmu}{\mbox{\boldmath $\mu$} {}}
\newcommand{\ddelta}{\mbox{\boldmath $\delta$} {}}
\newcommand{\kkappa}{\mbox{\boldmath $\kappa$} {}}
\newcommand{\llambda}{\mbox{\boldmath $\lambda$} {}}
\newcommand{\pomega}{\mbox{\boldmath $\varpi$} {}}
\newcommand{\ssigma}{\mbox{\boldmath $\sigma$} {}}
%
%
\newcommand{\DDDD}{\mbox{\boldmath ${\sf D}$} {}}
\newcommand{\IIII}{\mbox{\boldmath ${\sf I}$} {}}
\newcommand{\LLLL}{\mbox{\boldmath ${\sf L}$} {}}
\newcommand{\MMMM}{\mbox{\boldmath ${\sf M}$} {}}
\newcommand{\NNNN}{\mbox{\boldmath ${\sf N}$} {}}
\newcommand{\PPPP}{\mbox{\boldmath ${\sf P}$} {}}
\newcommand{\QQQQ}{\mbox{\boldmath ${\sf Q}$} {}}
\newcommand{\RRRR}{\mbox{\boldmath ${\sf R}$} {}}
\newcommand{\SSSS}{\mbox{\boldmath ${\sf S}$} {}}
\newcommand{\BBBBB}{\mbox{\boldmath ${\sf B}$} {}}
\newcommand{\tAAAA}{\tilde{\mbox{\boldmath ${\sf A}$}} {}}
\newcommand{\tDDDD}{\tilde{\mbox{\boldmath ${\sf D}$}} {}}
\newcommand{\tRRRR}{\tilde{\mbox{\boldmath ${\sf R}$}} {}}
\newcommand{\tQQQQ}{\tilde{\mbox{\boldmath ${\sf Q}$}} {}}
\newcommand{\AAAA}{\mbox{\boldmath ${\cal A}$} {}}
\newcommand{\BBB}{\mbox{\boldmath ${\cal B}$} {}}
\newcommand{\EMF}{\mbox{\boldmath ${\cal E}$} {}}
\newcommand{\GGG}{\mbox{\boldmath ${\cal G}$} {}}
\newcommand{\HHH}{\mbox{\boldmath ${\cal H}$} {}}
\newcommand{\QQQ}{\mbox{\boldmath ${\cal Q}$} {}}
\newcommand{\GGGG}{{\bf G} {}}
%
%
\newcommand{\ii}{{\rm i}}
\newcommand{\erf}{{\rm erf}}
\newcommand{\grad}{{\rm grad} \, {}}
\newcommand{\curl}{{\rm curl} \, {}}
\newcommand{\dive}{{\rm div}  \, {}}
\newcommand{\Dive}{{\rm Div}  \, {}}
\newcommand{\diag}{{\rm diag}  \, {}}
\newcommand{\sgn}{{\rm sgn}  \, {}}
\newcommand{\DD}{{\rm D} {}}
\newcommand{\DDD}{{\cal D} {}}
\newcommand{\dd}{{\rm d} {}}
\newcommand{\dV}{\,{\rm d}V {}}
\newcommand{\dS}{\,{\rm d}{{\bm{S}}} {}}
\newcommand{\const}{{\rm const}  {}}
\newcommand{\CR}{{\rm CR}}
\def\degr{\hbox{$^\circ$}}
\def\la{\mathrel{\mathchoice {\vcenter{\offinterlineskip\halign{\hfil
$\displaystyle##$\hfil\cr<\cr\sim\cr}}}
{\vcenter{\offinterlineskip\halign{\hfil$\textstyle##$\hfil\cr<\cr\sim\cr}}}
{\vcenter{\offinterlineskip\halign{\hfil$\scriptstyle##$\hfil\cr<\cr\sim\cr}}}
{\vcenter{\offinterlineskip\halign{\hfil$\scriptscriptstyle##$\hfil\cr<\cr\sim\cr}}}}}
\def\ga{\mathrel{\mathchoice {\vcenter{\offinterlineskip\halign{\hfil
$\displaystyle##$\hfil\cr>\cr\sim\cr}}}
{\vcenter{\offinterlineskip\halign{\hfil$\textstyle##$\hfil\cr>\cr\sim\cr}}}
{\vcenter{\offinterlineskip\halign{\hfil$\scriptstyle##$\hfil\cr>\cr\sim\cr}}}
{\vcenter{\offinterlineskip\halign{\hfil$\scriptscriptstyle##$\hfil\cr>\cr\sim\cr}}}}}
%
%
\def\Ta{\mbox{\rm Ta}}
\def\Ra{\mbox{\rm Ra}}
\def\Ma{\mbox{\rm Ma}}
\def\Co{\mbox{\rm Co}}
\def\Sh{\mbox{\rm Sh}}
\def\St{\mbox{\rm St}}
\def\Roo{\mbox{\rm Ro}^{-1}}
\def\Rooo{\mbox{\rm Ro}^{-2}}
\def\Pra{\mbox{\rm Pr}}
\def\Pran{\mbox{\rm Pr}}
\def\Sc{\mbox{\rm Sc}}
\def\Tr{\mbox{\rm Tr}}
\def\Rem{\mbox{\rm Rm}}
\def\Pm{{\rm Pm}}
\def\Rm{R_{\rm m}}
\def\Rmc{R_{\rm m,{\rm crit}}}
\def\Rey{\mbox{\rm Re}}
\def\Pe{\mbox{\rm Pe}}
\def\Co{\mbox{\rm Co}}
\def\Lu{\mbox{\rm Lu}}
\def\qp{q_{\rm p}}
\def\qs{q_{\rm s}}
\def\qe{q_{\rm e}}
\def\qi{q_{\rm i}}
\def\csz{c_{\rm s0}}
\def\cs{c_{\rm s}}
\def\pt{p_{\rm t}}
\def\ptz{p_{\rm t0}}
\def\vA{v_{\rm A}}
\def\hf{h_{\rm f}}
\def\hm{h_{\rm m}}
\def\kmean{k_{\rm m}}
\def\lf{l_{\rm f}}
\def\kd{k_{\rm d}}
\def\kf{k_{\rm f}}
\def\ellf{\ell_{\rm f}}
\def\tauf{\tau_{\rm f}}
\def\Ff{F_{\rm f}}
\def\Fm{F_{\rm m}}
\def\Hf{H_{\rm f}}
\def\Hm{H_{\rm m}}
\def\vArms{v_{\rm A,rms}}
\def\Brms{B_{\rm rms}}
\def\Urms{U_{\rm rms}}
\def\urms{u_{\rm rms}}
\def\uref{u_{\rm ref}}
\def\kappaBB{\kappa_{\rm BB}}
\def\kappaB{\kappa_{\rm B}}
\def\kappaOO{\kappa_{\Omega\Omega}}
\def\kappaO{\kappa_{\Omega}}
\def\kappah{\kappa_{\rm h}}
\def\kappat{\kappa_{\rm t}}
\def\kappatz{\kappa_{\rm t0}}
\def\nut{\nu_{\rm t}}
\def\etatz{\eta_{\rm t0}}
\def\nutz{\nu_{\rm t0}}
\def\etat{\eta_{\rm t}}
\def\mut{\mu_{\rm t}}
\def\muT{\mu_{\rm T}}
\def\uT{\mu{\rm T}}
\def\BBeq{|\BB|/B_{\rm eq}}
\def\Beq{B_{\rm eq}}
\def\cst{c_{\rm s}}
\newcommand{\ea}{{\rm et al.\ }}
\newcommand{\eaa}{{\rm et al.\ }}
\def\half{{\textstyle{1\over2}}}
\def\threehalf{{\textstyle{3\over2}}}
\def\threequarter{{\textstyle{3\over4}}}
\def\sevenhalf{{\textstyle{7\over2}}}
\def\onethird{{\textstyle{1\over3}}}
\def\onesixth{{\textstyle{1\over6}}}
\def\twothird{{\textstyle{2\over3}}}
\def\fourthird{{\textstyle{4\over3}}}
\def\quarter{{\textstyle{1\over4}}}
\newcommand{\W}{\,{\rm W}}
\newcommand{\V}{\,{\rm V}}
\newcommand{\kV}{\,{\rm kV}}
\newcommand{\T}{\,{\rm T}}
\newcommand{\G}{\,{\rm G}}
\newcommand{\Hz}{\,{\rm Hz}}
\newcommand{\nHz}{\,{\rm nHz}}
\newcommand{\kHz}{\,{\rm kHz}}
\newcommand{\kG}{\,{\rm kG}}
\newcommand{\K}{\,{\rm K}}
\newcommand{\g}{\,{\rm g}}
\newcommand{\s}{\,{\rm s}}
\newcommand{\mpers}{\,{\rm m/s}}
\newcommand{\ks}{\,{\rm ks}}
\newcommand{\cm}{\,{\rm cm}}
\newcommand{\cmcube}{\,{\rm cm^{-3}}}
\newcommand{\m}{\,{\rm m}}
\newcommand{\km}{\,{\rm km}}
\newcommand{\msec}{\,{\rm ms}}
\newcommand{\cms}{\,{\rm cm/s}}
\newcommand{\kms}{\,{\rm km/s}}
\newcommand{\gpercc}{\,{\rm g/cm}^3}
\newcommand{\kg}{\,{\rm kg}}
\newcommand{\ug}{\,\mu{\rm g}}
\newcommand{\kW}{\,{\rm kW}}
\newcommand{\MW}{\,{\rm MW}}
\newcommand{\Mm}{\,{\rm Mm}}
\newcommand{\Mx}{\,{\rm Mx}}
\newcommand{\pc}{\,{\rm pc}}
\newcommand{\kpc}{\,{\rm kpc}}
\newcommand{\yr}{\,{\rm yr}}
\newcommand{\Myr}{\,{\rm Myr}}
\newcommand{\Gyr}{\,{\rm Gyr}}
\newcommand{\erg}{\,{\rm erg}}
\newcommand{\mol}{\,{\rm mol}}
\newcommand{\dyn}{\,{\rm dyn}}
\newcommand{\J}{\,{\rm J}}
\newcommand{\RM}{\,{\rm RM}}
\newcommand{\EM}{\,{\rm EM}}
\newcommand{\AU}{\,{\rm AU}}
\newcommand{\A}{\,{\rm A}}
%
%
\newcommand{\yastroph}[2]{ #1, astro-ph/#2}
\newcommand{\ycsf}[3]{ #1, {Chaos, Solitons \& Fractals,} {#2}, #3}
\newcommand{\yepl}[3]{ #1, {Europhys.\ Lett.,} {#2}, #3}
\newcommand{\yaj}[3]{ #1, {AJ,} {#2}, #3}
\newcommand{\yjgr}[3]{ #1, {J.\ Geophys.\ Res.,} {#2}, #3}
\newcommand{\ysol}[3]{ #1, {Sol.\ Phys.,} {#2}, #3}
\newcommand{\yapj}[3]{ #1, {ApJ,} {#2}, #3}
\newcommand{\ypasp}[3]{ #1, {PASP,} {#2}, #3}
\newcommand{\yapjl}[3]{ #1, {ApJ,} {#2}, #3}
\newcommand{\yapjs}[3]{ #1, {ApJS,} {#2}, #3}
\newcommand{\yija}[3]{ #1, {Int.\ J.\ Astrobiol.,} {#2}, #3}
\newcommand{\yan}[3]{ #1, {Astron.\ Nachr.,} {#2}, #3}
\newcommand{\yzfa}[3]{ #1, {Z.\ f.\ Ap.,} {#2}, #3}
\newcommand{\ymhdn}[3]{ #1, {Magnetohydrodyn.} {#2}, #3}
\newcommand{\yana}[3]{ #1, {A\&A,} {#2}, #3}
\newcommand{\yanas}[3]{ #1, {A\&AS,} {#2}, #3}
\newcommand{\yanar}[3]{ #1, {A\&A Rev.,} {#2}, #3}
\newcommand{\yass}[3]{ #1, {Ap\&SS,} {#2}, #3}
\newcommand{\ygafd}[3]{ #1, {Geophys.\ Astrophys.\ Fluid Dyn.,} {#2}, #3}
\newcommand{\ygrl}[3]{ #1, {Geophys.\ Res.\ Lett.,} {#2}, #3}
\newcommand{\ypasj}[3]{ #1, {Publ.\ Astron.\ Soc.\ Japan,} {#2}, #3}
\newcommand{\yjfm}[3]{ #1, {J.\ Fluid Mech.,} {#2}, #3}
\newcommand{\ypepi}[3]{ #1, {Phys.\ Earth Planet.\ Int.,} {#2}, #3}
\newcommand{\ypf}[3]{ #1, {Phys.\ Fluids,} {#2}, #3}
\newcommand{\ypfb}[3]{ #1, {Phys.\ Fluids B,} {#2}, #3}
\newcommand{\ypp}[3]{ #1, {Phys.\ Plasmas,} {#2}, #3}
\newcommand{\ysov}[3]{ #1, {Sov.\ Astron.,} {#2}, #3}
\newcommand{\ysovl}[3]{ #1, {Sov.\ Astron.\ Lett.,} {#2}, #3}
\newcommand{\yjetp}[3]{ #1, {Sov.\ Phys.\ JETP,} {#2}, #3}
\newcommand{\yphy}[3]{ #1, {Physica,} {#2}, #3}
\newcommand{\yaraa}[3]{ #1, {ARA\&A,} {#2}, #3}
\newcommand{\yanf}[3]{ #1, {Ann. Rev. Fluid Mech.,} {#2}, #3}
\newcommand{\yprs}[3]{ #1, {Proc.\ Roy.\ Soc.\ Lond.,} {#2}, #3}
\newcommand{\yprt}[3]{ #1, {Phys.\ Rep.,} {#2}, #3}
\newcommand{\yprl}[3]{ #1, {Phys.\ Rev.\ Lett.,} {#2}, #3}
\newcommand{\yphl}[3]{ #1, {Phys.\ Lett.,} {#2}, #3}
\newcommand{\yptrs}[3]{ #1, {Phil.\ Trans.\ Roy.\ Soc.,} {#2}, #3}
\newcommand{\ymag}[3]{ #1, {Magnetohydrodynamics,} {#2}, #3}
\newcommand{\ymn}[3]{ #1, {MNRAS,} {#2}, #3}
\newcommand{\ynat}[3]{ #1, {Nature,} {#2}, #3}
\newcommand{\yptrsa}[3]{ #1, {Phil. Trans. Roy. Soc. London A,} {#2}, #3}
\newcommand{\yphm}[3]{ #1, {Phil. Mag.,} {#2}, #3}
\newcommand{\ypria}[3]{ #1, {Proc. R. Irish Acad.,} {#2}, #3}
\newcommand{\ysci}[3]{ #1, {Science,} {#2}, #3}
\newcommand{\ysph}[3]{ #1, {Solar Phys.,} {#2}, #3}
\newcommand{\yspj}[3]{ #1, {Sov. Phys. JETP,} {#2}, #3}
\newcommand{\ypr}[3]{ #1, {Phys.\ Rev.,} {#2}, #3}
\newcommand{\ypre}[3]{ #1, {Phys.\ Rev.\ E,} {#2}, #3}
\newcommand{\ypnas}[3]{ #1, {Proc.\ Nat.\ Acad.\ Sci.,} {#2}, #3}
\newcommand{\yicarus}[3]{ #1, {Icarus,} {#2}, #3}
\newcommand{\yspd}[3]{ #1, {Sov.\ Phys.\ Dokl.,} {#2}, #3}
\newcommand{\yjcp}[3]{ #1, {J.\ Comput.\ Phys.,} {#2}, #3}
\newcommand{\yjour}[4]{ #1, {#2}, {#3}, #4}
\newcommand{\yprep}[2]{ #1, {\sf #2}}
\newcommand{\ybook}[3]{ #1, {#2} (#3)}
\newcommand{\yproc}[5]{ #1, in {#3}, ed.\ #4 (#5), #2}
\newcommand{\pproc}[4]{ #1, in {#2}, ed.\ #3 (#4), (in press)}
\newcommand{\pprocc}[5]{ #1, in {#2}, ed.\ #3 (#4, #5)}
\newcommand{\pmn}[1]{ #1, {MNRAS}, to be published}
\newcommand{\pana}[1]{ #1, {A\&A}, to be published}
\newcommand{\papj}[1]{ #1, {ApJ}, to be published}
\newcommand{\ppapj}[3]{ #1, {ApJ}, {#2}, to be published in the #3 issue}
\newcommand{\sprl}[1]{ #1, {PRL}, submitted}
\newcommand{\sapj}[1]{ #1, {ApJ}, submitted}
\newcommand{\sana}[1]{ #1, {A\&A}, submitted}
\newcommand{\smn}[1]{ #1, {MNRAS}, submitted}

\title{Pumping velocity in homogeneous helical turbulence with shear}

\author{Igor Rogachevskii}
\email{gary@bgu.ac.il}
\author{Nathan Kleeorin}
\email{nat@bgu.ac.il}
\affiliation{Department of Mechanical Engineering,
Ben-Gurion University of the Negev, P.O.Box 653, Beer-Sheva 84105,  Israel}
\author{Petri \ J.\ K\"apyl\"a}
\email{petri.kapyla@helsinki.fi}
\affiliation{Department of Physics, Gustaf H\"allstr\"omin katu
2a (PO Box 64), FI-00064 University of Helsinki, Finland}
\affiliation{NORDITA, AlbaNova University Center,
Roslagstullsbacken 23, SE-10691 Stockholm, Sweden}
\author{Axel Brandenburg}
\email{brandenb@nordita.org}
\affiliation{NORDITA, AlbaNova University Center,
Roslagstullsbacken 23, SE-10691 Stockholm, Sweden}
\affiliation{Department of Astronomy, Stockholm University, SE 10691 Stockholm, Sweden}

\date{\today,~ $ $Revision: 1.154 $ $}

\begin{abstract}
Using different analytical methods (the quasi-linear approach,
the path-integral technique and tau-relaxation approximation)
we develop a comprehensive mean-field theory for a pumping
effect of the mean magnetic field in homogeneous
non-rotating helical turbulence with imposed large-scale shear.
The effective pumping velocity
is proportional to the product of $\alpha$
effect and large-scale vorticity associated with the shear,
and causes a separation of the toroidal
and poloidal components of the mean magnetic field
along the direction of the mean vorticity.
We also perform direct numerical simulations of sheared
turbulence in different ranges of hydrodynamic and
magnetic Reynolds numbers and
use a kinematic test-field method to determine
the effective pumping velocity.
The results of the numerical simulations are in agreement
with the theoretical predictions.
\end{abstract}


\maketitle

\section{Introduction}

The origin of cosmic magnetic fields is one of the
fundamental problems in theoretical physics and astrophysics.
It is generally believed that solar and galactic
magnetic fields are caused by the combined action of helical
turbulent motions of fluid and differential rotation
\cite{M78,P79,KR80,ZRS83,RSS88,O03,BS05}.
In most of these studies, differential rotation plays
merely the role of enhancing the magnetic field in the
toroidal direction.
However, in recent years there has been increased interest in mean-field
effects caused specifically by turbulent shear flows.
This interest is caused by discoveries of the shear dynamo \cite{RK03,RK04}
and vorticity dynamo \cite{EKR03,EGKR07}
in non-helical homogeneous turbulence with a large-scale shear.
In particular, recent numerical experiments
\cite{BR05,YHS08,YHR08,BRRK08,KKB08,HP09} have clearly demonstrated
the existence of a shear dynamo of a large-scale magnetic field in
non-helical turbulence or turbulent convection with
superimposed large-scale shear.
However, the origin of the shear dynamo is still subject of
active discussions
\cite{RK03,RK04,RS06,RK06,BRRK08,KR08,SS09,SS10,SKR08}.

There are three additional phenomena that are also related
to the presence of shear.
One is the vorticity dynamo, which is the self-excitation of large-scale
vorticity in a turbulence with large-scale shear.
It has been predicted theoretically \cite{EKR03,EGKR07}
and detected in recent numerical experiments
\cite{YHS08,YHR08,KMB09}.
The vorticity dynamo can also affect the dynamo process
of the mean magnetic field.
Another phenomenon is a non-zero $\alpha$ effect in
non-helical turbulence with shear when the system
is inhomogeneous or density stratified.
In that case there is an $\alpha$ effect \cite{RK03,RS06}
that can lead to an alpha-shear dynamo.
Finally, when homogeneous turbulence with shear is helical,
there is an effective pumping velocity $\bm{\gamma}
\propto \alpha {\bm W}$ of the large-scale magnetic field,
where ${\bm W}$ is the large-scale vorticity caused by shear.
This effect has so far only been found in direct
numerical simulations (DNS) \cite{MKB09},
but there has so far been no theory for this new effect,
nor has there been a systematic survey of DNS for
determining the dependence of pumping on magnetic
Reynolds and Prandtl numbers
as well as the turbulent Mach number.

The goal of the present study is to develop a
comprehensive theory of mean-field pumping in homogeneous
helical turbulence with shear
and to perform systematic numerical simulations
designed for detailed comparison with the
theoretical predictions.
It is important to emphasize that the pumping of the
large-scale magnetic field
discussed usually in the literature has always been
connected with inhomogeneous turbulence
\cite{KR80,Kit91,RKR03}, but here we study
the pumping for homogeneous, albeit helical turbulence.

\section{Governing equations}

We consider homogeneous helical turbulence with a linear
shear velocity $\meanUU = (0, Sx, 0)$.
Averaging the induction equation over an ensemble
of turbulent velocity field yields the
mean-field equation:
\begin{eqnarray}
{\partial \overline{\bm B} \over \partial t} &=&
\bec{\nabla} {\rm \times} \left(\overline{\bm U}
{\bm \times} \overline{\bm B} + \overline{{\bm u}
{\bm \times} {\bm b}} - \eta \bec{\nabla}
{\bm \times} \overline{\bm B} \right),
\label{L1}
\end{eqnarray}
where ${\cal E}_i\equiv(\overline{{\bm u} \times {\bm b}})_i
=a_{ij} \overline{B}_j + b_{ijk} \nabla_k \overline{B}_j$
is the mean electromotive force,
${\bm u}$ and ${\bm b}$ are the fluctuations of velocity
and magnetic field, overbars denote averaging
over an ensemble of turbulent velocity fields,
$\meanBB$ is the mean magnetic field,
$\meanUU$ is the mean velocity that includes
only the imposed large-scale shear, and
$\eta$ is the magnetic diffusion due to electrical
conductivity of the fluid.
Note that the part $a_{ij} \overline{B}_j$ in the expression for
the mean electromotive force determines the effective pumping
velocity, $\gamma_i=-{1\over2}\epsilon_{ijk}a_{ij}$,
and the $\alpha$ tensor, $\alpha_{ij}={1\over2}(a_{ij}+a_{ji})$, i.e.,
${\cal E}_i^{(a)} = \alpha_{ij} \meanB_j + (\bm{\gamma}
\times \meanBB)_i$, while the turbulent magnetic diffusion
and the shear-current dynamo effect are associated with
the $b_{ijk}$ term.

To determine the turbulent transport coefficients
in homogeneous helical turbulence with mean velocity
shear we use the following equations for fluctuations of velocity
and magnetic field:
\begin{eqnarray}
{\partial {\bm u} \over \partial t} &=& - (\meanUU {\bm \cdot}
\nab) {\bm u} - ({\bm u} {\bm \cdot} \nab) \meanUU -
{1 \over \meanR} \, \nab p + {1 \over 4 \pi \meanR} \,
\big[({\bm b} {\bm \cdot} \nab) \meanBB
\nonumber\\
& & + (\meanBB {\bm \cdot} \nab){\bm b}\big] + \nu \Delta {\bm u}
+ {\bm u}^{N} + {\bm f}^{(u)} ,
\label{A1} \\
{\partial {\bm b} \over \partial t} &=& (\meanBB {\bm \cdot}
\nab){\bm u} - ({\bm u} {\bm \cdot} \nab) \meanBB
+ ({\bm b} {\bm \cdot} \nab)\meanUU - (\meanUU {\bm \cdot}
\nab) {\bm b}
\nonumber\\
& &  + \eta \, \Delta {\bm b} + {\bm b}^N ,
\label{A2}
\end{eqnarray}
where $\nu$ is the kinematic viscosity,
$\meanR$ is the mean density
of the incompressible fluid flow,
$p$ is the fluctuation of total
(hydrodynamic and magnetic) pressure, the magnetic
permeability of the fluid is included in the definition
of the magnetic field, ${\bm v}^{N}$ and ${\bm b}^{N}$
are the nonlinear terms, and $\meanR \, {\bm f}^{(u)}$ is the
stirring force for the background velocity fluctuations.

We begin by deriving expressions for the pumping effect
that are valid in different regimes, where fluid
and magnetic Reynolds numbers are both small, both are large,
or only the fluid Reynolds number is large, but the magnetic
Reynolds number is small.
These results will then be compared with those of DNS in
the corresponding regimes.

\subsection{Small magnetic and hydrodynamic Reynolds numbers}

We use the quasi-linear or second order correlation
approximation (SOCA) applied to shear flow turbulence
(see \cite{RS06,KR08}).
This approach is valid for small magnetic and hydrodynamic
Reynolds numbers. To exclude the pressure term
from the equation of motion~(\ref{A1}) we
calculate $\nab {\bm \times} (\nab {\bm \times} {\bm
u})$, then we rewrite the obtained equation and Eq.~(\ref{A2})
in Fourier space, apply the two-scale approach (i.e., we use
large-scale and small-scale variables),
neglect nonlinear terms in Eqs.~(\ref{A1})--(\ref{A2}),
but retain molecular dissipative terms in
these equations. We seek a solution for fluctuations
of velocity and magnetic fields as an expansion for weak
velocity shear:
\begin{eqnarray}
{\bm u} &=& {\bm u}^{(0)} + {\bm u}^{(1)} + ... \;,
\label{C4} \\
{\bm b} &=& {\bm b}^{(0)} + {\bm b}^{(1)} + ... \;,
\label{C5}
\end{eqnarray}
where
\begin{eqnarray}
&& b_i^{(0)}({\bm k}, \omega) = G_\eta(k, \omega)
\, \biggl[i({\bm k} {\bm \cdot} \meanBB) \delta_{ij}
- \Big(\delta_{ij} \, k_{m} {\partial \over \partial k_{n}}
\nonumber\\
&& \quad + \delta_{im} \delta_{jn}\Big) (\nabla_{n} \meanB_{m})
\biggr] \, u_j^{(0)}({\bm k}, \omega) ,
 \label{C1}\\
&& u_i^{(1)}({\bm k}, \omega) = G_\nu(k, \omega) \, \biggl(
2 k_{iq} \delta_{jp} + \delta_{ij} \, k_{q} {\partial \over
\partial k_{p}} - \delta_{iq} \delta_{jp} \biggr)
\nonumber\\
&& \quad \times (\nabla_{p} \meanU_{q}) \, u_j^{(0)}({\bm k}, \omega) ,
\label{C2}\\
&& b_i^{(1)}({\bm k}, \omega) = G_\eta(k, \omega) \, \biggl\{
\Big[i({\bm k} {\bm \cdot} \meanBB) \delta_{ij}
- \Big(\delta_{ij} \, k_{m} {\partial \over \partial k_{n}}
\nonumber\\
&& \quad + \delta_{im} \delta_{jn} \Big)
\, (\nabla_{n} \meanB_{m}) \Big] \, u_j^{(1)}({\bm k}, \omega)
+ \Big[\delta_{ij} \, k_{q} {\partial \over \partial k_{p}}
\nonumber\\
&& \quad + \delta_{iq} \delta_{jp}\Big] \,
b_j^{(0)}({\bm k}, \omega)\, (\nabla_{p} \meanU_{q}) \biggr\}.
\label{C3}
\end{eqnarray}
Here $G_\nu(k, \omega) = (\nu k^2 - i \omega)^{-1}$,
$\, G_\eta(k, \omega) = (\eta k^2 - i \omega)^{-1}$,
and $\delta_{ij}$ is the Kronecker tensor.
The statistical properties of the
background velocity fluctuations with a zero
large-scale shear, $\uu^{(0)}$, are assumed to be given.
For derivation of Eqs.~(\ref{C1})--(\ref{C3}) we
use the identity
\begin{eqnarray*}
\int \meanU_q({\bm Q}) \, b_n({\bm k} - {\bm Q})
\,d{\bm Q} = i (\nabla_{p} \meanU_{q}) \,
{\partial b_n \over \partial k_{p}} ,
\end{eqnarray*}
which is valid in the framework of the mean-field approach,
i.e., it is assumed that there is scale separation.
Equations~(\ref{C1})--(\ref{C3}) coincide with those derived by
\cite{RS06}, and they allow us to determine the
cross-helicity tensor
$g_{ij}^{(1)} = \langle u_i^{(0)} \, b_j^{(1)}  \rangle
+ \langle u_i^{(1)} \, b_j^{(0)} \rangle$.
This procedure yields the
contributions ${\cal E}_{m}^{(S)} = \varepsilon_{mij} \,
\int \, g_{ij}^{(1)}({\bm k}, \omega) \,d {\bm k} \,d \omega$
to the mean electromotive force caused by sheared helical turbulence.
We are interested first of all in the contributions to the mean electromotive
force which are proportional to the mean magnetic field, i.e.,
${\cal E}_i^{(a)} = \alpha_{ij} \meanB_j + (\bm{\gamma}
\times \meanBB)_i$.
For the integration in $\omega$-space and in
${\bm k}$-space we have to specify
a model for the background shear-free helical turbulence (with
$\meanBB = 0)$, which is determined by equation:
\begin{eqnarray}
&& \langle u_i({\bm k},\omega) \, u_j(-{\bm k},-\omega)
\rangle^{(0)} =  {E(k) \, \Phi(\omega) \over 8 \pi \, k^{2}}
\,  \Big[\Big(\delta_{ij} - {k_i\,k_j\over k^2}
\Big)
\nonumber\\
&& \quad \times \langle {\bm u}^2 \rangle^{(0)}
- {i \over k^2} \, \varepsilon_{ijl} \, k_l \, \langle{\bm u}
\cdot (\nab \times {\bm u}) \rangle^{(0)}
\Big],
 \label{AB1}
\end{eqnarray}
where $E(k)$ is the energy spectrum (e.g., a power-law
spectrum, $E(k) \propto (k/\kf)^{-q}$ with the exponent $1<q<3$
for the wavenumbers $\kf \leq k \leq \kd$, where $\kf$ and $\kd$ are
the forcing and dissipation wavenumbers), and $\varepsilon_{ijk}$
is the fully antisymmetric Levi-Civit\`{a} tensor.
We consider  the frequency function $\Phi(\omega)$
in the form of the Lorentz profile:
$\Phi(\omega)=\nu k^2 / [\pi \,(\omega^2 + \nu^2 k^4)]$.
This model for the frequency function
corresponds to the correlation function
\begin{eqnarray}
\langle u_i(t) u_j(t+\tau) \rangle \propto
\exp (-\tau \, \nu k^2).
\end{eqnarray}
In that case, and under the assumption of small
magnetic and hydrodynamic Reynolds
numbers, the effective pumping velocity,
$\bm{\gamma}$, and the off-diagonal components of the
tensor $\alpha_{ij}$  are given by
\begin{eqnarray}
\bm{\gamma} &=&  {C_1(q) \over 2} \, \left({{\rm Pm} \over
1+ {\rm Pm}}\right)^2
\, {\rm Re}^2 \, \, \tauf \, \alpha_\ast \, \meanWW,
\label{new-vel-small-Rm}\\
\alpha_{ij}&=& {C_1(q) \over 5} \, {(2 {\rm Pm} + 1) \, {\rm Pm}
\over (1+ {\rm Pm})^2} \, {\rm Re}^2
\, \tauf \, \alpha_\ast \, (\partial \meanU)_{ij},
\label{new-alpha-small-Rm}\\
C_1(q) &=& \int_{\kf}^{\kd} E(k) \, \left({k\over \kf}\right)^{-4} \,dk
\nonumber\\
&=& \left({q-1\over q+3}\right) \,
\left[{1 - (\kf/\kd)^{q+3} \over 1 - (\kf/\kd)^{q-1}}\right] ,
\end{eqnarray}
where $\alpha_\ast = - (\tauf/3) \, \langle{\bm u} \cdot
(\nab \times {\bm u}) \rangle^{(0)}$,
$\, {\rm Pm}=\nu/\eta$ is the magnetic Prandtl number,
${\rm Re}=\tauf \, \langle {\bm u}^2 \rangle^{(0)} / \nu$
is the hydrodynamic Reynolds number, ${\rm Rm}= {\rm Re}
\, {\rm Pm}$ is the
magnetic Reynolds number, $\tauf = \ellf / u_{\rm rms}$ is the turnover time,
where $\, \ellf = 1 / \kf $ is the energy-containing (forcing)
scale of a random velocity field, and $u_{\rm rms}=
\sqrt{\langle{\bm u}^2\rangle^{(0)}}$.
For the integration in $\omega$-space we use the integrals $I_n(k)$ given in \App{Append-A}.
For linear shear velocity, $\meanUU = (0, Sx, 0)$, the mean vorticity
is $\meanWW = \nab {\bm \times} \meanUU =(0,0,S)$,
and the mean symmetric tensor $(\partial \meanU)_{ij} =
(\nabla_i \meanU_{j} + \nabla_j \meanU_{i}) / 2$
has only two nonzero components: $(\partial \meanU)_{12}
=(\partial \meanU)_{21} = S/2$.
Therefore, $\alpha_{ij}$ has two non-zero off-diagonal components
caused by both, shear and helical turbulence $\alpha_{12}=\alpha_{21}$,
while the effective pumping velocity, $\bm{\gamma}$, has only
one component directed along the vertical axis, $\bm{\gamma}
= (0, 0, \gamma)$:
\begin{eqnarray}
\gamma &=&  {C_1(q) \over 2} \, \left({{\rm Pm} \over
1+ {\rm Pm}}\right)^2 \, {\rm Re}^2 \, \,\alpha_\ast \,\, {\rm Sh},
\label{v-small-Rm}\\
\alpha_{_{12}}&=& \alpha_{_{21}}= {C_1(q) \over 10} \,
{(2 {\rm Pm} + 1) \, {\rm Pm} \over (1+ {\rm Pm})^2}
\, {\rm Re}^2 \, \alpha_\ast \,\, {\rm Sh},
\label{a-small-Rm}
\end{eqnarray}
where ${\rm Sh}= \tauf \, S$ is the shear parameter.
As follows from
Eqs.~(\ref{v-small-Rm}) and (\ref{a-small-Rm}),
$\gamma \propto {\rm Pm}^2$ and $\alpha_{_{12}}
\propto {\rm Pm}$ for ${\rm Pm} \ll 1$, while for
${\rm Pm} \gg 1$ the effective pumping velocity $\gamma$
and $\alpha_{_{12}}$ are independent of ${\rm Pm}$.
For all values of the magnetic
Prandtl numbers, $\gamma$ and $\alpha_{12}$ are positive.
This asymptotic behavior which is valid for
${\rm Re} \ll 1$, is in agreement with Figs.~\ref{pgam_pm}
and~\ref{paij_pm} (see Sect. III).
Note that the diagonal components of the tensor $\alpha_{ij}$
in this case are
\begin{eqnarray}
\alpha &=& - {C_2(q) \over 3} \, \left({{\rm Rm} \over
1+ {\rm Pm}}\right) \, \tauf \, \langle{\bm u} \cdot
(\nab \times {\bm u}) \rangle^{(0)} ,
\nonumber\\
\label{al1}\\
C_2(q) &=& \int_{\kf}^{\kd} E(k) \, \left({k\over \kf}\right)^{-2} \,dk
\nonumber\\
&=& \left({q-1\over q+1}\right) \,
\left[{1 - (\kf/\kd)^{q+1} \over 1 - (\kf/\kd)^{q-1}}\right] .
\label{al2}
\end{eqnarray}

\subsection{Large magnetic and hydrodynamic Reynolds numbers}

To determine the
the effective pumping velocity and the tensor $\alpha_{ij}$
in homogeneous helical turbulence with mean velocity
shear for large magnetic and hydrodynamic Reynolds numbers
we use the procedure which is similar to that
applied in \cite{RK04} in
earlier investigations of shear flow turbulence.
Let us derive equations for the second moments.
We apply the two-scale approach, e.g., we use large
scale ${\bm R} = ( {\bm x} +
{\bm y}) / 2$, $\, {\bm K} = {\bm k}_1 + {\bm k}_2$
and small scale ${\bm r} = {\bm x} - {\bm y}$,
$\, {\bm k} = ({\bm k}_1 - {\bm k}_2)
/ 2$ variables (see, e.g., \cite{RS75}).
We derive equations for the following correlation functions:
\begin{eqnarray*}
f_{ij}({\bm k}) &=& \hat L(u_i; u_j) , \; h_{ij}({\bm k}) =
\hat L(b_i; b_j) ,
\\
g_{ij}({\bm k}) &=& (4 \pi \meanR)^{-1} \, \hat L(b_i; u_j) ,
\end{eqnarray*}
where
\begin{eqnarray*}
\hat L(a; c) = \int \langle a({\bm k} + {\bm  K} / 2)
c(-{\bm k} +{\bm  K} / 2) \rangle
\, \exp{(i {\bm K} {\bm \cdot} {\bm R}) } \,d {\bm  K} ,
\end{eqnarray*}
and $\langle ... \rangle$ denotes averaging over ensemble
of turbulent velocity field.
The equations for these correlation
functions are given by (see \cite{RK04})
\begin{eqnarray}
{\partial f_{ij}({\bm k}) \over \partial t} \!&=&\! i({\bm k} {\bm
\cdot} \meanBB) \Phi_{ij} + I^f_{ij} + I_{ijmn}^S(\meanUU)
f_{mn}+ F_{ij}
+ \hat{\cal N} f_{ij},
\nonumber \\
{\partial h_{ij}({\bm k}) \over \partial t} &=& - i({\bm k}{\bm
\cdot} \meanBB) \Phi_{ij} + I^h_{ij} + E_{ijmn}^S(\meanUU)
h_{mn} + \hat{\cal N} h_{ij} ,
\nonumber \\
{\partial g_{ij}({\bm k }) \over \partial t} &=& i({\bm k} {\bm
\cdot} \meanBB) [f_{ij}({\bm k}) - h_{ij}({\bm k})
- h_{ij}^{(H)}] +I^g_{ij}
\nonumber \\
&& + J_{ijmn}^S(\meanUU) g_{mn} + \hat{\cal N} g_{ij} ,
\label{B8}
\end{eqnarray}
where hereafter we omit the arguments
$t$ and ${\bm R}$ in the
correlation functions and neglect small terms
$ \sim O(\nabla^2)$.
Here $F_{ij}$ is related to the forcing term and
$\nab = \partial / \partial {\bm R} $.
In Eqs.~(\ref{B8}),
$\Phi_{ij}({\bm k}) = (4 \pi \meanR)^{-1} \, [g_{ij}({\bm k}) -
g_{ji}(-{\bm k})]$, and $ \hat{\cal N} f_{ij}$,
$\, \hat{\cal N}h_{ij}$,
$\, \hat{\cal N}g_{ij}$, are the third-order moments
appearing due to the nonlinear terms which
include also molecular dissipation terms.
The tensors $I_{ijmn}^S(\meanUU)$,
$\, E_{ijmn}^S(\meanUU)$
and $J_{ijmn}^S(\meanUU)$ are given by
\begin{eqnarray*}
I_{ijmn}^S(\meanUU) &=& \biggl(2 k_{iq}
\delta_{mp} \delta_{jn}
+ 2 k_{jq} \delta_{im} \delta_{pn}
- \delta_{im} \delta_{jq}
\delta_{pn}
\nonumber\\
&& - \delta_{iq} \delta_{jn} \delta_{pm}
+ \delta_{im} \delta_{jn}
k_{q} {\partial \over \partial k_{p}} \biggr)
\nabla_{p} \meanU_{q} ,
\nonumber\\
E_{ijmn}^S(\meanUU) &=& \biggl(\delta_{im} \delta_{jq}
\delta_{pn} + \delta_{jm} \delta_{iq} \delta_{pn}
\nonumber\\
& & + \delta_{im} \delta_{jn} k_{q} {\partial
\over \partial k_{p}}
\biggr) \nabla_{p} \meanU_{q} ,
\nonumber\\
J_{ijmn}^S(\meanUU) &=& \biggl(2 k_{jq}
\delta_{im} \delta_{pn}
- \delta_{im} \delta_{pn} \delta_{jq}
+ \delta_{jn} \delta_{pm}
\delta_{iq}
\nonumber\\
& & + \delta_{im} \delta_{jn} k_{q} {\partial
\over \partial k_{p}}
\biggr) \nabla_{p} \meanU_{q} ,
\end{eqnarray*}
where $k_{ij} = k_i k_j / k^2$.
The source terms $I_{ij}^f$,
$I_{ij}^h$, and $I_{ij}^g$ which contain
the large-scale spatial derivatives
of the magnetic field $\meanBB$, are
given in \cite{RK04}. Next, in
Eqs.~(\ref{B8}) we split the tensor for
magnetic fluctuations into nonhelical, $h_{ij},$
and helical, $h_{ij}^{(H)},$ parts.
The helical part of the tensor of magnetic
fluctuations $h_{ij}^{(H)}$ depends on the magnetic
helicity and it follows from magnetic helicity
conservation arguments
(see, e.g., \cite{KR82,GD94,GD95,KR99} and \cite{BS05} for a review).

The second-moment equations include the first-order
spatial differential operators $\hat{\cal N}$
applied to the third-order moments $M^{(\rm III)}$.
A problem arises how to close the system, i.e.,
how to express the set of the third-order terms
$\hat{\cal N} M^{(\rm III)}$ through the lower moments
$M^{(\rm II)}$ (see, e.g., \cite{O70,MY75,Mc90}).
We use the spectral $\tau$-closure-approximation
which postulates that the deviations of the
third-moment terms,
$\hat{\cal N} M^{(\rm III)}({\bm k})$, from the
contributions to these terms due to
the background turbulence, $\hat{\cal N}
M^{(\rm III,0)}({\bm k})$, are expressed through
the similar deviations of the second moments,
$M^{(\rm II)}({\bm k}) - M^{(\rm II,0)}({\bm k})$:
\begin{eqnarray}
\hat{\cal N} M^{(\rm III)}({\bm k}) &-& \hat{\cal N}
M^{(\rm III,0)}({\bm k})
\nonumber\\
&=& - {1 \over \tau_r(k)} \, \Big[M^{(\rm II)}({\bm k})
- M^{(\rm II,0)}({\bm k})\Big],
\label{A3}
\end{eqnarray}
(see, e.g., \cite{O70,PFL76,KRR90}), where $\tau_r(k)$
is the scale-dependent relaxation time, which can
be identified with the correlation time, $\tauf$,
of the turbulent velocity field for large
hydrodynamic and magnetic Reynolds numbers.
The quantities with the superscript $(0)$ correspond
to the background shear-free turbulence with
a zero mean magnetic field.
We apply the spectral $\tau$ approximation only
for the nonhelical part $h_{ij}$
of the tensor of magnetic fluctuations.
Note that a justification of the $\tau$ approximation
for different situations has been performed
in a number of numerical simulations and analytical
studies
(see, e.g., \cite{BF02,BF03,FB02,BK04,BS05,BSM05,BS07,SSB07,RK07}).

We take into account that the characteristic time of variation of
the magnetic field $\meanBB$ is substantially longer than the
correlation time $\tauf$. This allows us
to obtain a stationary solution for Eqs.~(\ref{B8}) for the
second-order moments, $M^{(\rm II)}({\bm k})$, which are the sums of
contributions caused by shear-free and
sheared turbulence. The contributions to the mean
electromotive force caused by a shear-free turbulence
and sheared non-helical turbulence are
given in \cite{RK04}. In particular, the contributions to the
electromotive force caused by the sheared
turbulence read:
${\cal E}_{m}^{(S)} = \varepsilon_{mji} \, \int \,
g_{ij}^{(S)}({\bm k})
\,d {\bm k} $, where the corresponding contributions to the
cross-helicity
tensor $g_{ij}^{(S)}$ in the kinematic approximation, are given by
\begin{eqnarray}
g_{ij}^{(S)}({\bm k}) &=& i \tau_r(k) \, \Big[J_{ijmn}^S \,
\tau_r(k) \, ({\bm k} {\bm \cdot} \meanBB)
\nonumber\\
&& \,
+ \tau_r(k) \, ({\bm k} {\bm \cdot} \meanBB) \, I_{ijmn}^S \Big]
\, f_{mn}^{(0)} ,
\label{B2}
\end{eqnarray}
and we use the following model
for the background shear-free helical turbulence (with $\meanBB = 0)$:
\begin{eqnarray}
&& f_{ij}^{(0)}=\langle u_i({\bm k}) \, u_j(-{\bm k},) \rangle^{(0)}
=  \Big[ \Big(\delta_{ij} - {k_i \, k_j \over k^2}\Big)
\, \langle {\bm u}^2 \rangle^{(0)}
\nonumber\\
&& \quad \quad - {i \over k^2} \, \varepsilon_{ijl} \, k_l \,
\langle{\bm u} \,{\bm \cdot} \,(\nab {\bm \times} \, {\bm u})
\rangle^{(0)} \Big] \, {E(k) \over 8 \pi \, k^{2}},
 \label{AB2}
\end{eqnarray}
where the energy
spectrum is $E(k) = (q-1) \, (k / k_{\rm f})^{-q} ,$ $\, k_{\rm f} = 1 /
\ell_{\rm f}$ and the length $\, \ell_{\rm f}$ is the maximum scale of turbulent
motions. The turbulent correlation time is $\tau_r(k) = 2 \, \tauf \,
(k / k_{\rm f})^{1-q}$.
Therefore, for large magnetic and hydrodynamic Reynolds number
the effective pumping velocity,
$\bm{\gamma}$, and the off-diagonal components of the tensor
$\alpha_{ij}$
caused by sheared helical turbulence
are given by
\begin{eqnarray}
&& \bm{\gamma} = {2 \over 3} \, \tauf \, \alpha_\ast \, \meanWW,
\label{vel-large-Rm}\\
&& \alpha_{ij}= - {4 \over 5} \, (5- 2 q)  \,
\tauf \, \alpha_\ast \, (\partial \meanU)_{ij} .
\label{alpha-large-Rm}
\end{eqnarray}
Since the mean symmetric tensor $(\partial \meanU)_{ij}$
has only two nonzero components: $(\partial \meanU)_{12}
=(\partial \meanU)_{21} = S/2$, the tensor
$\alpha_{ij}$ has only two non-zero off-diagonal components,
$\alpha_{12}=\alpha_{21}$. In particular,
\begin{eqnarray}
&& \gamma = {2 \over 3} \, \alpha_\ast \, {\rm Sh} ,
\label{v-large-Rm}\\
&& \alpha_{12}=\alpha_{21}= - {2 \over 5} \, (5- 2 q)  \,
\alpha_\ast \, {\rm Sh} = - {2 \over 3} \, \alpha_\ast \, {\rm Sh} ,
\label{a-large-Rm}
\end{eqnarray}
where we have used the Kolmogorov
kinetic energy spectrum exponent $q=5/3$ in Eq.~(\ref{a-large-Rm}).
The diagonal components of the tensor $\alpha_{ij}$
in this case are $\alpha=\alpha_\ast$
(see, e.g., \cite{KR80,M78}).
These results for large magnetic and hydrodynamic Reynolds number
are in qualitative agreement with DNS performed
in \cite{MKB09}.

\subsection{Large magnetic Reynolds numbers and small hydrodynamic
Reynolds numbers}

To develop a mean-field theory for large magnetic Reynolds
numbers and small hydrodynamic Reynolds numbers we use stochastic
calculus for a random velocity field.
To derive an equation for the mean magnetic field we use an
exact solution of the induction
equation for the total field ${\bm B}$
(which is the sum of the mean $\meanBB$
and fluctuating ${\bm b}$ parts)
with an initial condition
${\bm B}(t=t_0,{\bm x}) = {\bm B}(t_0,{\bm x})$ in the form of
a functional integral:
\begin{eqnarray}
B_{i}(t,{\bm x}) = \langle G_{ij}(t,t_0,\bec{\xi}) \,
\exp(\bec{\hat \xi} \cdot \nab) B_{j}(t_0,{\bm x}) \rangle_{\bm w},
\label{M1}
\end{eqnarray}
(see, e.g., \cite{DM84,KRS02}), where the operator $\exp(\bec{\hat \xi} \cdot \nab)$
is determined by
\begin{eqnarray}
\exp(\bec{\hat \xi} \cdot \nab) &=& \sum_{k=0}^{\infty} \, {1 \over k!} \,
(\bec{\hat \xi} \cdot \nab)^{k} \;,
\label{DD7}
\end{eqnarray}
$\bec{\hat \xi} = \bec{\xi} - {\bm x}$ (see \App{Append-B}).
The Wiener trajectory $\bec{\xi}(t,t_0,{\bm x})$ is determined by
\begin{eqnarray}
\bec{\xi}(t,t_0,{\bm x}) = {\bm x} - \int_{0}^{t-t_0} {\bm v}(t_\sigma,\bec{\xi})
\,d\sigma + (2 \eta)^{1/2} {\bm w}(t-t_0) ,
\nonumber\\
\label{M2}
\end{eqnarray}
where $t_\sigma = t - \sigma$, and the velocity field ${\bm v}$ is the sum
of the mean shear velocity $\meanUU$ and fluctuating ${\bm u}$ parts.
We consider large magnetic Reynolds number, but take into account
small yet finite magnetic diffusion $\eta$.
The magnetic diffusion can be described by
a random Wiener process
${\bm w}(t)$ that is defined by the following properties:
$\langle w_i(t) \rangle_{\bm w} = 0$ and $\langle w_i(t+\tau)
w_j(t)\rangle_{\bm w} = \tau \delta_{ij}$, where $\langle \cdot
\rangle_{\bm w}$ denotes the averaging over the
statistics of the Wiener random process.
The function $G_{ij}(t,s, \bec{\xi})$ is determined by equation:
\begin{eqnarray}
{d G_{ij}(t,s,\bec{\xi}) \over ds} = N_{ik} G_{kj}(t,s,\bec{\xi}) ,
\label{M3}
\end{eqnarray}
with the initial condition $G_{ij}(t=s) =\delta_{ij}$ and
$N_{ij} = \nabla_j v_{i}$.
The form of the exact solution~(\ref{M1})
allows us to separate the averaging over random Brownian motion
of particles (i.e., the averaging over a random Wiener process
${\bm w}(t)$) and a random velocity ${\bm u}$.

We consider a random flow with a small yet finite
Strouhal number (that is the ratio the correlation time
of a random fluid flow to the turnover time $\ell_f / u_{\rm rms}$).
A random velocity field with a small Strouhal number
can be modelled by a random velocity field
with a constant renewal time $\tau$.
Assume that in the intervals $\ldots (- \tau, 0); (0,
\tau); (\tau, 2 \tau); \ldots $ the velocity fields are
statistically independent and have the same statistics.
This implies that the velocity field looses memory at the prescribed
instants $t = m \tau$, where $m = 0, \pm 1, \pm 2, \ldots$.
This velocity field cannot be considered as a stationary  velocity
field for small times $\sim \tau$, however, it behaves like a
stationary field for $t \gg \tau$.
Averaging Eq.~(\ref{M1}) over the random velocity field
we arrive at  the equation for the mean magnetic field,
$\meanBB$:
\begin{eqnarray}
{\partial \meanB_{i} \over \partial t} &=& \big[\nab \times
(\meanUU {\bf \times} \meanBB)\big]_i + A_{ijm} \nabla_m \meanB_{j}
\nonumber\\
&& + D_{ijmn} \nabla_m \nabla_n \meanB_{j} ,
\label{M5}
\end{eqnarray}
(see \App{Append-B}), where
\begin{eqnarray}
A_{ijm} &=& {1 \over \tau} \langle \langle \hat \xi_m \, G_{ij}
\rangle \rangle_{\bm w} \;,
\label{M6}\\
D_{ijmn} &=& {1 \over 2 \tau} \langle \langle \hat \xi_m \hat
\xi_n \, G_{ij}  \rangle \rangle_{\bm w} ,
\label{M7}
\end{eqnarray}
the angular brackets $\langle \cdot \rangle$ denote an ensemble
average over the random velocity field.
Therefore, the mean magnetic field is determined by double averaging
over two independent random processes, i.e., by the ensemble average
over the random velocity field and by the average over Wiener random
process ${\bm w}(t)$.

We are interested in the lowest-order contributions to the mean electromotive
force which are proportional to the mean magnetic field,
${\cal E}_{i}^{(a)} = a_{ij} \, \overline{B}_j$, where $a_{ij}= (1/2)
\varepsilon_{inm} \, A_{njm}$ and the tensor $A_{ijm}$ reads:
\begin{eqnarray}
A_{ijm} &=& -{1 \over \tau} \int_{0}^{\tau} d t \, \int_{0}^{\tau} d t'
\, \left\langle \big[v_{m}(t,\bec{\xi})\big]_{\bf x}  \, \big[\nabla_j v_{i}(t',\bec{\xi})\big]_{\bf y}\right\rangle  ,
\nonumber\\
\label{M8}
\end{eqnarray}
where ${\bf x} \to {\bf y}$ and $\big[v_{m}(t,\bec{\xi})\big]_{\bf x}$
denotes the Eulerian velocity determined at the Wiener trajectory
$\bec{\xi}$ that passes through the point ${\bf x}$ at instant $t$.
Hereafter the angular brackets denote double averaging over a
random velocity field and over the statistics of the Wiener process.

For small hydrodynamic Reynolds numbers we seek the solutions of
the linearized Navier-Stokes equation~(\ref{A1}) for incompressible
velocity field ${\bm u}$ as superpositions of the Orr-Kelvin random
shearing waves ${\bm u}(t,{\bm r}) = \int {\bm u}(t,{\bm k}_0) \exp
[i {\bm k}(t) \cdot {\bm r}] \, d{\bm k}_0$, where
${\bm k}_0=(k_{x0}, k_y, k_z)$, $\, {\bm k}(t)=
(k_{x0}-S k_y t,k_y,k_z)$ (see, e.g., \cite{K1887,Oa1907,Ob1907,SKR08}).
Therefore, the effective pumping velocity,
$\bm{\gamma}$, and the off-diagonal components of the tensor
$\alpha_{ij}$ are given by
\begin{eqnarray}
\gamma_n &=& {1 \over 2} \, \varepsilon_{nji} \, a_{ij} =
{1 \over 4} \, A_{kmm} = - {i \over 4 \tau} \int_{0}^{\tau} \,d t
\, \int_{0}^{\tau} \, d t'
\nonumber\\
&& \times \, k_{m}(t') \, \langle v_{m}(t,{\bm k}_0) \,
v_{n}^*(t',{\bm k}_0) \rangle,
\label{M9}\\
\alpha_{ij} &=& {1 \over 2} \, (a_{ij} + a_{ji}) =
{1 \over 4} \, \left(\varepsilon_{inm} \, A_{njm} +
\varepsilon_{jnm} \, A_{nim} \right)
\nonumber\\
&=& - {i \over 4 \tau} \int_{0}^{\tau} \,d t \,
\int_{0}^{\tau} \, d t'
\, \left(\varepsilon_{inm} \, k_{j}(t') + \varepsilon_{jnm}
\,  k_{i}(t') \right)
\nonumber\\
&& \times \langle v_{m}(t,{\bm k}_0) \, v_{n}^*(t',{\bm k}_0) \rangle .
\label{M10}
\end{eqnarray}
Using these equations
and Eqs.~(\ref{L8})--(\ref{L10}) in \App{Append-C} we obtain the
effective pumping velocity, $\bm{\gamma}=(0, 0, \gamma)$,
and the off-diagonal components
$\alpha_{12}=\alpha_{21}$ of the tensor $\alpha_{ij}$ for
large magnetic Reynolds numbers and small hydrodynamic Reynolds numbers:
\begin{eqnarray}
&& \gamma = {C_1(q) \over 3} \, \alpha_\ast \, {\rm Sh} \, {\rm Re}^2,
\label{M11}\\
&& \alpha_{12}=\alpha_{21} = \left(C_2(q) \,  {\rm Re} \,
{\tau \over \tauf} - {3 \, C_1(q) \over 2}  \, {\rm Re}^2
\right)  \, \alpha_\ast \, {\rm Sh} ,
\nonumber\\
\label{M12}
\end{eqnarray}
where ${\rm Re} \ll \tau/\tauf < 1$.
The diagonal components of the tensor $\alpha_{ij}$
in this case obtained using path-integral approach
are $\alpha= - (1/3) \, \langle \tau {\bm u} \cdot
(\nab \times {\bm u}) \rangle^{(0)}$
(see, e.g., \cite{DM84,RK97}).
In the next section we discuss comparison
with new systematic DNS designed for comparison
with our theoretical predictions.

\section{Comparison with DNS}

\subsection{Numerical model}
Our DNS model is identical to that used in \cite{MKB09}.
We begin by testing the analytical results numerically using
three-dimensional simulations of isotropically forced turbulence in a
fully periodic cube of size $(2\pi)^3$. The uniform shear
$\meanUU=(0,Sx,0)$ is imposed using
the shearing box method and the gas obeys an isothermal equation of
state characterized by the constant speed of sound $\cst$. We solve
the continuity and Navier--Stokes equations in the form
\begin{equation}
\frac{{\mathcal D} \ln \rho}{{\mathcal D} t} =
    - {\bm U} \cdot \bm{\nabla} \ln \rho - \bm\nabla \cdot {\bm U},
\label{dlnrho_dt}
\end{equation}
\begin{equation}
\frac{{\mathcal D} {\bm U}}{{\mathcal D} t} =
    - {\bm U} \cdot \bm{\nabla} {\bm U} -
S U_x \bm{\hat{y}} - \cs^2\bm{\nabla} \ln \rho + {\bm f} + {\bm F}_{{\rm visc}},
\label{duu_dt}
\end{equation}
where the imposed shear is subsumed in the
advective derivative
\begin{equation}
\frac{{\mathcal D}}{{\mathcal D} t}\equiv \frac{\pd}{\pd t} + Sx\frac{\pd}{\pd y}.
\end{equation}
Here $\rho$ is the density, ${\bm U}$ is the velocity, ${\bm f}$
describes the forcing, and ${\bm F}_{\rm visc}= \rho^{-1}\bm\nabla
\cdot (2\rho \nu \mbox{\boldmath ${\sf S}$})$ is the viscous force, where $\nu$
is the kinematic viscosity, and
\begin{equation}
{\sf S}_{ij} = \half (U_{i,j} + U_{j,i}) - \onethird \bm\nabla \cdot \bm{U}
\end{equation}
is the traceless rate of strain tensor.
The forcing function $\bm f$ is given in \cite{B01}:
\begin{equation}
\ff(\xx,t)={\rm Re}\{N\ff_{\kk(t)}\exp[\ii\kk(t)\cdot\xx+\ii\phi(t)]\},
\end{equation}
where $\xx$ is the position vector. The wavevector $\kk(t)$ and the
random phase $-\pi<\phi(t)\le\pi$ change at every time step, so
$\ff(\xx,t)$ is $\delta$-correlated in time. The normalization factor
$N$ is chosen on dimensional grounds to be $N=f_0 c_{\rm
  s}(|\kk|c_{\rm s}/\delta t)^{1/2}$, where $f_0$ is a nondimensional
forcing amplitude. At each timestep we select randomly one of many
possible wavevectors in a certain range around a given forcing
wavenumber. The average wavenumber is referred to as $k_{\rm f}$. In
the present study we always use $\kef/k_1=5$. We force the system with
transverse helical waves \cite{HBD04},
\begin{equation}
\ff_{\kk}=\RRRR\cdot\ff_{\kk}^{\rm(nohel)}\quad\mbox{with}\quad
{\sf R}_{ij}={\delta_{ij}-\ii\sigma\epsilon_{ijk}\hat{k}_k
\over\sqrt{1+\sigma^2}},
\end{equation}
where $\sigma=1$ for the fully helical case with
positive helicity of the forcing function,
\EQ
\ff_{\kk}^{\rm(nohel)}=
\left(\kk\times\eee\right)/\sqrt{\kk^2-(\kk\cdot\eee)^2},
\label{nohel_forcing}
\EN
is a non-helical forcing function, and $\eee$ is an arbitrary unit vector
not aligned with $\kk$; note that $|\ff_{\kk}|^2=1$.
We use fully helical forcing, i.e.\ $\sigma=1$, in all of our runs.

The boundary conditions in the $y$ and $z$ directions are periodic,
whereas shearing-periodic conditions are used in the $x$ direction.
The simulations are governed by the fluid and magnetic Reynolds
numbers, the magnetic Prandtl number, and the shear and Mach numbers:
\begin{eqnarray}
\Rey&=&\frac{\urms}{\nu\kef},\,\,\,
\Rem=\frac{\urms}{\eta\kef},\,\,\,
\Pm=\frac{\nu}{\eta},\,\,\,\\
\nonumber
\Sh&=&\frac{S}{\urms\kef},\,\,\,
\Ma=\frac{\urms}{\cst}.
\end{eqnarray}
Here $\urms$ is the root mean square velocity of turbulent
motions and $\eta$ is the magnetic diffusivity.
We use the {\sc Pencil
Code}\footnote{http://pencil-code.googlecode.com/} to perform the
simulations.

\subsection{Test field method}
We apply the kinematic test-field method
(see, e.g., \cite{Sch05,Sch07,BRRK08}) to compute
the effective pumping velocity, $\bm{\gamma}$,
and all components of the tensor $\alpha_{ij}$.
The essence of this method is that a set of prescribed
test fields $\meanBB^{(p,q)}$ and the flow from the DNS are used to
evolve separate realizations of small-scale fields ${\bm b}^{(p,q)}$.
Neither the test
fields $\meanBB^{(p,q)}$ nor the small-scale fields ${\bm b}^{(p,q)}$
act back on the flow. These small-scale fields are then used to
compute the electromotive force $\meanEMF^{(p,q)}$ corresponding to
the test field $\meanBB^{(p,q)}$. The number and form of the test
fields used depends on the problem at hand. For the purposes of the
present study we use uniform horizontal test fields
$\meanBB^{(1)}=(B_0,0,0)$ and $\meanBB^{(2)}=(0,B_0,0)$, in which case
the series expansion of the electromotive force contains only a single
term
\begin{equation}
\mathcal{E}_i^{(a)}=a_{ij}\mean{B}_j.
\end{equation}
We present the results using the quantities:
\begin{eqnarray}
\alpha & = & \half(a_{11}+a_{22}),\\
\alpha_{12} & = & \alpha_{21}=\half(a_{21}+a_{12}),\\
\gamma & = & \half(a_{21}-a_{12}).
\end{eqnarray}
We use $\alpha_0=\onethird \urms$ as a normalization factor when
presenting numerical results.
Errors are estimated by dividing the time series into three equally
long parts and computing time averages for each of them.
The largest departure from the time average computed over
the entire time series represents the error.
This definition of the error bar gives an indication
about the mean value that one would obtain for shorter
parts of the time series.
With this definition, the error bars do normally become shorter
for longer runs, provided the time series is stationary.
This would not be the case for the rms value of the deviations,
which might sometimes also be of interest.

\begin{table*}
\caption{Summary of the runs.
}
\vspace{12pt}
\centerline{\begin{tabular}{lccccc}
    Set & $\Rey$ & $\Pm$ & $\Sh$ & $\Ma$ & grid \\
    \hline
    A1   & 0.04 & 0.05\ldots25 & $-0.20$ & $0.010$ & $32^3\ldots64^3$   \\
    A2   & 0.16 & 0.02\ldots20 & $-0.13$ & $0.016$ & $32^3\ldots64^3$  \\
    \hline
    B1   & 0.08\ldots81 & 1 & $-0.025$ & $0.080$ & $32^3\ldots256^3$   \\
    B2   & 0.08\ldots83 & 1 & $-0.075$ & $0.080$ & $32^3\ldots256^3$  \\
    B3   & 0.08\ldots3.5 & 1 & $-0.25$ & $0.080$ & $32^3$   \\
    B4   & 0.08\ldots0.4 & 1 & $-2.5$ & $0.080$ & $32^3$  \\
    \hline
    C1   & 0.04 & 1 & $-0.020\ldots-0.19$ & $0.010$ & $32^3$   \\
    C2   & 0.16 & 1 & $-0.012\ldots-0.12$ & $0.016$ & $32^3$   \\
    C3   & 0.45 & 1 & $-0.009\ldots-0.09$ & $0.023$ & $32^3$   \\
    C4   & 1.3  & 1 & $-0.006\ldots-0.07$ & $0.032$ & $32^3$   \\
    \hline
    D1   & 0.08 & 1 & $-0.010$ & $0.002\ldots0.41$ & $32^3$   \\
    \hline
\label{runs}\end{tabular}}\end{table*}

\begin{table}
\caption{
Convergence study of
$\gamma$ and $\alpha_{21}$ for $\Rem=1.3$ and $\Sh=-0.06$ from
simulations with different grid sizes.
  }
\vspace{12pt}
\centerline{\begin{tabular}{lccc}
    Run & $\gamma/\alpha_0 [10^{-2}]$ & $\alpha_{21}/\alpha_0 [10^{-2}]$ & grid \\
    \hline
    E1   & $1.02\pm0.12$ & $0.94\pm0.25$ & $16^3$  \\
    E2   & $1.05\pm0.07$ & $0.89\pm0.20$ & $32^3$  \\
    E3   & $0.99\pm0.06$ & $0.83\pm0.55$ & $64^3$  \\
    E4   & $0.94\pm0.18$ & $0.88\pm0.23$ & $128^3$ \\
    \hline
\label{convergence}
\end{tabular}}
\end{table}

\subsection{Results}

We perform several sets of simulations where we vary the parameters
$\Pm$, $\Rem$, $\Sh$, and $\Ma$ individually to study the
analytical results derived in Sect.~II; see Table~\ref{runs}.
The setup used here is prone to
exhibit the so-called vorticity dynamo \citep{EKR03,KMB09}, due to
which large-scale vorticity is generated, and complications can arise
in the interpretation of the simulation data. Here we restrict the
studied parameter range so that the values of $\Rey$ and $\Sh$ are
subcritical for the vorticity dynamo.
In our runs where the Reynolds numbers are of the order of unity or
less, a low grid resolution of $32^3$ is often sufficient.
Indeed, in Table~II we show the results obtained for different
resolutions ranging from $16^3$ to $128^3$ for $\Rem$ around 1,
which demonstrates good convergence of the results within error bars.

\subsubsection{Dependence on $\Pm$}
\begin{figure}
\begin{center}
\includegraphics[width=\columnwidth]{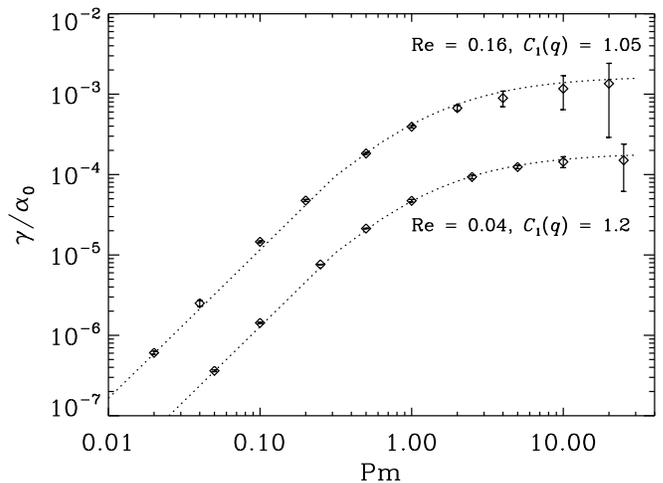}
\end{center}
\caption[]{Pumping coefficient $\gamma=\half(a_{21}-a_{12})$
  normalized by $\alpha_0=\onethird \urms$ as a function of $\Pm$
  for two values of ${\rm Re}$ (Sets~A1 and A2).
  The shear parameter $\Sh=-0.20$ ($-0.13$) for $\Rey=0.04$ ($0.16$).
  Analytical results according to Eq.~(\ref{v-small-Rm}) are
  overplotted with dotted lines. The values of $C_1(q)$ are used
  as fit parameters and indicated in the legends.}
\label{pgam_pm}
\end{figure}

Figure~\ref{pgam_pm} shows our results for $\gamma$ as a function of
magnetic Prandtl number $\Pm$. We find that the numerical results
coincide with the analytical formula, Eq.~(\ref{v-small-Rm}).
Values of the order of $C_1(q)\approx 1$ fit the DNS results within
the error estimates.

\begin{figure}
\begin{center}
\includegraphics[width=\columnwidth]{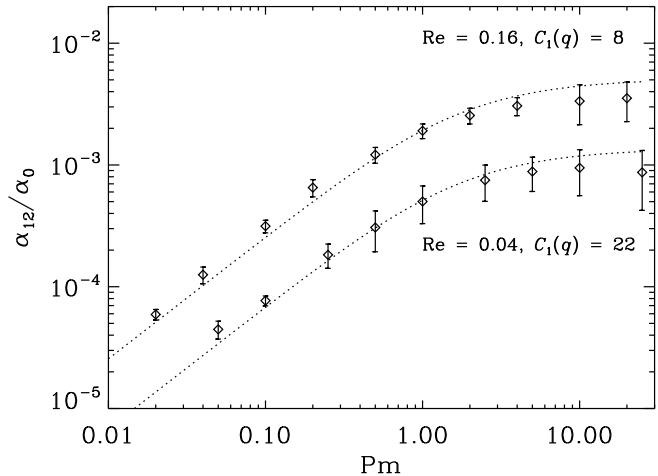}
\end{center}
\caption[]{Symmetric part of $a_{ij}$,
  $\alpha_{12}=\half(a_{21}+a_{12})$ normalized by
  $\alpha_0=\onethird \urms$ as a function of $\Pm$ for the same runs
  as in Fig.~\ref{pgam_pm}. The dotted lines show the analytical
  result according to Eq.~(\ref{a-small-Rm}), with the values of
  $C_1(q)$ indicated in the legends.}
\label{paij_pm}
\end{figure}

\begin{figure}
\begin{center}
\includegraphics[width=\columnwidth]{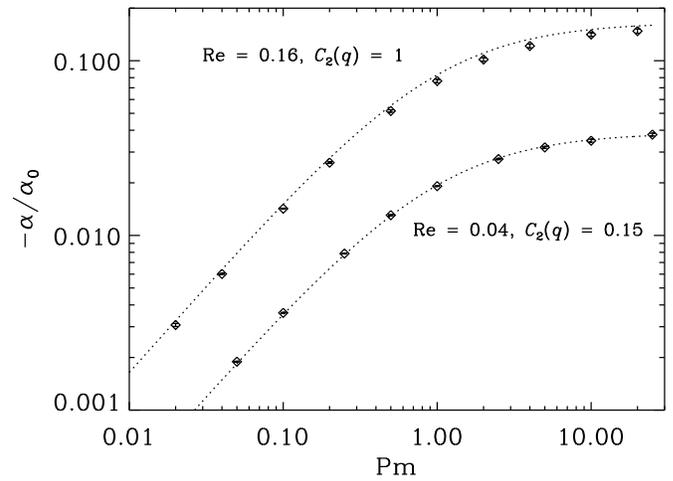}
\end{center}
\caption[]{$\alpha$-effect as a function of $\Pm$ normalized by
  $\alpha_0=\onethird \urms$ for the same runs
  as in Fig.~\ref{pgam_pm}.
  Analytical results according to Eq.~(\ref{al1}) are
  overplotted with dotted lines. The values of $C_2(q)$ are used as fit
  parameters and indicated in the legends.}
\label{palp_pm}
\end{figure}

Figure~\ref{paij_pm} shows the results for $\alpha_{12}$ as a function
of $\Pm$ for two values of $\Rey$. The data for $\alpha_{12}$ shows
significantly larger fluctuations than the corresponding results for
$\gamma$. However, the DNS results seem to fall in line with the
analytical expression, Eq.~(\ref{a-small-Rm}), although the value of
$C_1(q)$ needed to fit the data is an order of magnitude larger than in
the case of $\gamma$.
This can be explained by comparing Eqs.~(\ref{M11}) and~(\ref{M12}),
which show that $\gamma \propto {\rm Re}^2$, while
$\alpha_{12} \propto {\rm Re} \, (\tau / \tauf)$,
where $\tau$ is the flow renovating time,
and $\tauf=\ellf/u_{\rm rms}$ is the turnover time of
turbulent eddies.
Note that Eqs.~(\ref{M11}) and~(\ref{M12}) are obtained for large magnetic Reynolds numbers, while ${\rm Re} \ll \tau/\tauf < 1$.
This implies that for these conditions $\alpha_{12} \gg \gamma$.
The latter is in agreement with DNS results (see Figs.~\ref{pgam_pm}
and~\ref{paij_pm}).

In Fig.~\ref{palp_pm} we show $\alpha$-effect (the diagonal elements
In Fig.~\ref{palp_pm} we show the $\alpha$-effect (the diagonal elements
of the $\alpha_{ij}$ tensor) as a function of the magnetic Prandtl number
$\Pm$. These results are in a good agreement
with the analytical results~(\ref{al1}).

\begin{figure}
\begin{center}
\includegraphics[width=\columnwidth]{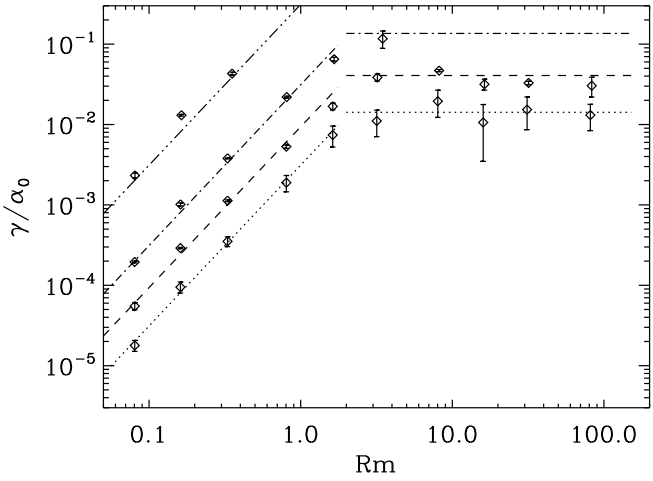}
\end{center}
\caption[]{$\gamma$ as a function of $\Rem$ for $\Pm=1$ and
  for four values of $\Sh$ ($-0.025$, $-0.075$, $-0.25$,
  and $-2.5$; see Sets~B1 to B4).
  The lines show the analytical results according to
  Eqs.~(\ref{v-small-Rm}) and (\ref{v-large-Rm}) with $C_1(q)=1$, for
  Sets~B1 (dotted lines), B2 (dashed), B3 (dot-dashed), and B4
  (triple-dot dashed), respectively.}
\label{pgam_rm}
\end{figure}

\begin{figure}
\begin{center}
\includegraphics[width=\columnwidth]{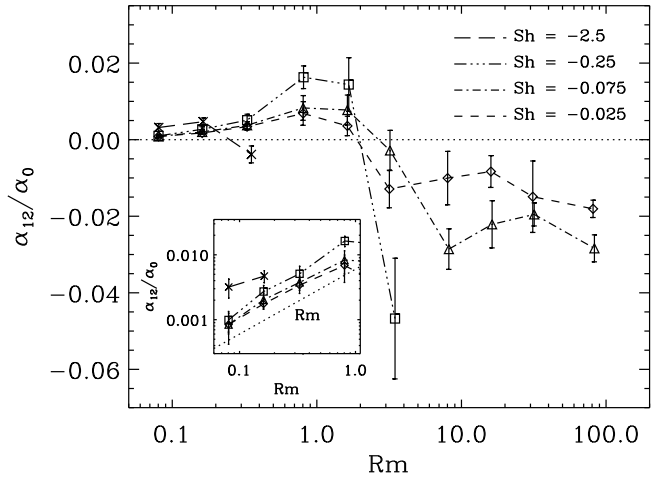}
\end{center}
\caption[]{Symmetric contribution $\alpha_{12}$ as a function
of $\Rem$ for $\Pm=1$ and four values of shear as indicated by the legend (Sets~B1 to B4).}
\label{paij_rm}
\end{figure}

\begin{figure}
\begin{center}
\includegraphics[width=\columnwidth]{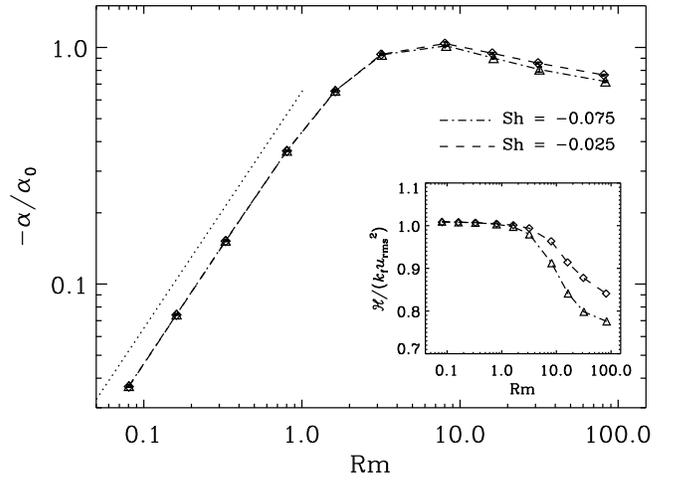}
\end{center}
\caption[]{$\alpha$-effect as a function of $\Rem$ normalized by
  $\alpha_0=\onethird \urms$ for two values of ${\rm Re}$ (Sets~B1 and B2).
  The dotted line is proportional to $\Rem$.
  The inset shows the normalized kinetic helicity of the flow.}
\label{palp_rm}
\end{figure}

\subsubsection{Dependence on $\Rem$}

Our results for $\gamma$ as a function of $\Rem$ are shown in
Fig.~\ref{pgam_rm}. We find that for $\Rem$ smaller than roughly two,
$\gamma$ is well described by the analytical result,
Eq.~(\ref{v-small-Rm})
obtained for $\Rem \ll 1$ and $\Rey \ll 1$.
For greater $\Rem$, $\gamma$ is consistent
with a constant value as a function of $\Rem$,
and is in accordance with Eq.~(\ref{v-large-Rm})
derived for $\Rem \gg 1$ and $\Rey \gg 1$.
Note also that for the largest values of the
shear parameter, $\Sh=-2.5$ ($-0.25$), there is a vorticity dynamo
for $\Rem>1$ ($\Rem>3$), so no points are plotted in those cases.

The off-diagonal component $\alpha_{12}$, shown in Fig.~\ref{paij_rm},
is proportional to $\Rey$ for small $\Rem$, while the analytical
expression~(\ref{a-small-Rm}) yields $\alpha_{12} \propto \Rey^2$.
A sign change occurs for $\Rem\approx 2$, and the values of $\alpha_{12}$
are consistently negative in this regime in agreement with
Eq.~(\ref{a-large-Rm}) derived for $\Rem \gg 1$ and $\Rey \gg 1$.
The data is noisy but suggest that $\alpha_{12}$ could be independent
of $\Rem$ at high $\Rem$ in an agreement with the analytical
result~(\ref{a-large-Rm}).
Furthermore, for small $\Rem$ the dependence on shear is weak,
although a clearer dependence on shear is seen for $\Rem$ greater
than around 10.

In Fig.~\ref{palp_rm} we show $\alpha$ as a function of $\Rem$.
We find that $\alpha$ is proportional to $\Rem$ for small
magnetic Reynolds numbers in agreement with Eq.~(\ref{al1}).
For $\Rem$ greater than roughly five,
$\alpha$ decreases slightly, while the theory suggests
that $\alpha$ is independent of $\Rem$ for $\Rem \gg 1$.
This inconsistency can be understood in terms of the
relative kinetic helicity $\mathcal{H}/(\kef \urms^2)$, where
$\mathcal{H}=\overline{\bm\omega\cdot{\bm u}}$, which
decreases by about 20 per cent between $\Rem$ 8 and 83
(see the inset in Fig.~\ref{palp_rm}).
Since $\alpha \propto \mathcal{H}$, this explains the decrease of
$\alpha$ with $\Rem$ for $\Rem \gg 1$.

\begin{figure}
\begin{center}
\includegraphics[width=\columnwidth]{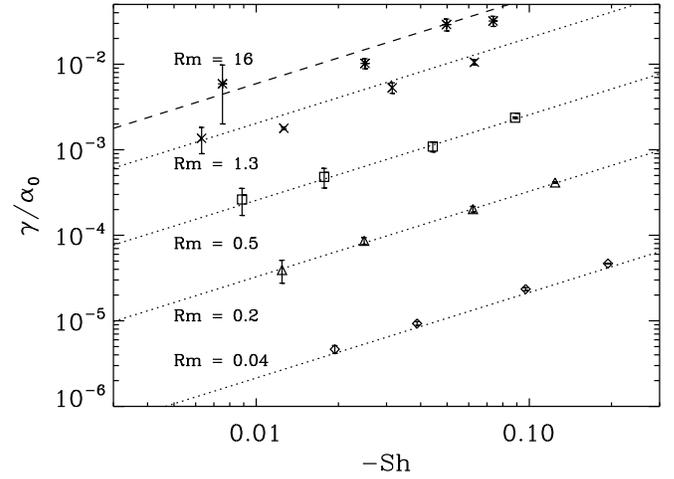}
\end{center}
\caption[]{Pumping velocity $\gamma=\half(a_{21}-a_{12})$
normalized by $\alpha_0$
as a function of $\Sh$ for $\Pm=1$ and different values of $\Rem$ as
indicated in the legend (Sets~C1--C4).
Analytical results according to Eqs.~(\ref{v-small-Rm}) with
$C_1(q)=1$, and (\ref{v-large-Rm}) are overplotted with dotted and
dashed lines, respectively.}
\label{pgam}
\end{figure}

\subsubsection{Dependence on shear}

\begin{figure}
\begin{center}
\includegraphics[width=\columnwidth]{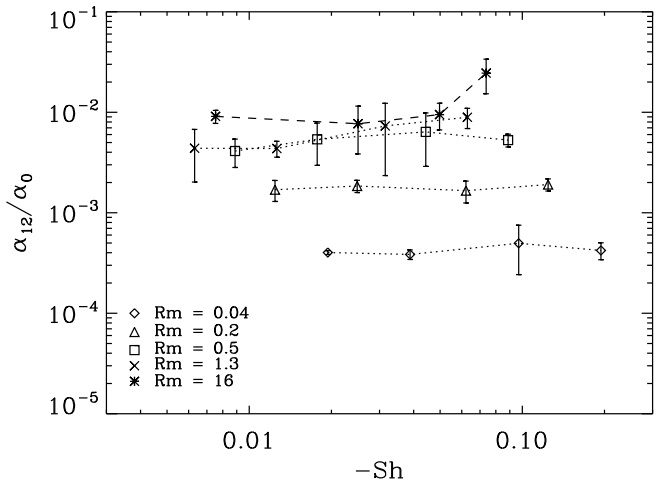}
\end{center}
\caption[]{Symmetric contribution $\alpha_{12}$ normalized
by $\alpha_0$ as a function of $\Sh$ for $\Pm=1$ and
different values of $\Rem$ as
indicated in the legend (Sets~C1--C4).
Runs with $\Rem=16$ are shown with asterisks and
  connected by a dashed line. For these runs $\alpha_{12}<0$ so the
  plot shows $-\alpha_{12}$.}
\label{paij2}
\end{figure}

Figure~\ref{pgam} shows the pumping velocity
$\gamma$ normalized by $\alpha_0$ as a function of
the shear number, $\Sh$, for $\Pm=1$ and
different values of $\Rem$.
Linear dependence of $\gamma$ on shear is clearly seen in
Fig.~\ref{pgam}.
This is in agreement with the analytical result of Eq.~(\ref{v-small-Rm}).
Rather surprisingly, the data for $\alpha_{12}$ suggest
that there is no dependence on shear
(Fig.~\ref{paij2}),
in contradiction with the analytical result
of Eq.~(\ref{a-small-Rm})
that was derived for small shear, $S \tauf \ll 1$.

\begin{figure}
\begin{center}
\includegraphics[width=\columnwidth]{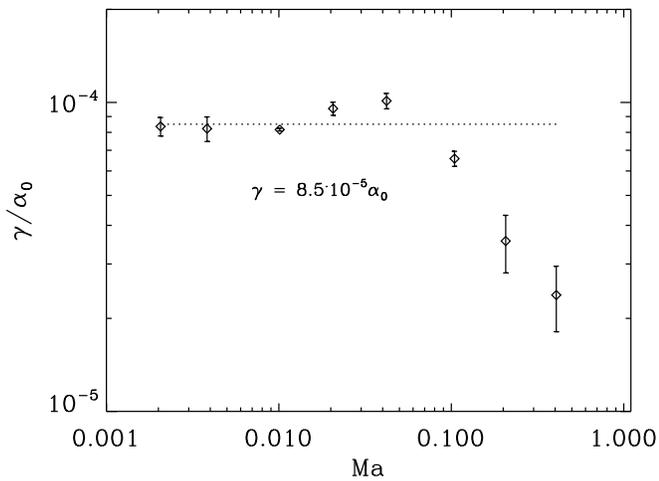}
\end{center}
\caption[]{Pumping coefficient $\gamma=\half(a_{21}-a_{12})$ as a
  function of the Mach number for ${\rm Pm}=1$ (Set~D1). The normalization
  factor is $\alpha_0=\onethird \urms$, and $\Sh=-0.10$.}
\label{pgam_Ma}
\end{figure}
Note that our theory has been developed for incompressible flow
since the DNS results are nearly independent of Mach number
for Ma $< 0.05$.
This is shown in Fig.~\ref{pgam_Ma}, where we notice a sharp decline
of $\gamma$ for larger values of the Mach number.
We are not aware of similar findings for mean-field transport coefficients as a function of Mach number.

\section{Discussion and Conclusions}
\label{Concl}

To clarify the physical effect related to the pumping
velocity, $\bm{\gamma}$, and the off-diagonal components
of the tensor $\alpha_{ij}$ we rewrite the contributions to the
mean electromotive force which are proportional to the mean
magnetic field in the following form:
\begin{eqnarray}
{\cal E}_i^{(S)} &=& \alpha_{ij} \meanB_j + (\bm{\gamma}
\times \meanBB)_i
\nonumber\\
&=& \left[\bm{\gamma}^{(P)} \times \meanBB^{(P)} + \bm{\gamma}^{(T)} \times \meanBB^{(T)}\right]_i,
\label{S1}
\end{eqnarray}
where $\meanBB^{(T)}$ is the toroidal mean magnetic field directed
along the mean shear velocity $\meanUU$ (along the $y$ axis),
$\meanBB^{(P)}$ is the poloidal mean magnetic field directed perpendicular
to both, the mean shear velocity $\meanUU$ and the mean vorticity
(along the $x$ axis), while the pumping velocities, $\bm{\gamma}^{(T)}$ and $\bm{\gamma}^{(P)}$, of the
toroidal and poloidal components of the mean magnetic field are given by:
\begin{eqnarray}
\bm{\gamma}^{(P)} &=& \hat{\bm z} \, (\alpha_{12} +\gamma) ,
\label{S2}\\
\bm{\gamma}^{(T)} &=& - \hat{\bm z} \, (\alpha_{12} - \gamma) .
\label{S3}
\end{eqnarray}
Here we take into account the following identities for the off-diagonal
components of the tensor $\alpha_{ij} = (\hat{x}_i \hat{y}_j
+ \hat{x}_j \hat{y}_i) \, \alpha_{12}$ and $\alpha_{ij} \meanB_j
= \alpha_{12} \, \hat{\bm z} \times (\meanBB^{(P)}-\meanBB^{(T)})$,
where $\alpha_{12}=\alpha_{21}$ and $\hat{\bm x}$, $\, \hat{\bm y}$,
$\, \hat{\bm z}$ are the unit vectors directed along $x$, $y$
and $z$ axes, respectively.

It follows from these equations that, when $\alpha_{12} > \gamma > 0$,
the effective pumping velocity of the poloidal mean magnetic field is
directed upward (along the $z$ axis), while the effective pumping velocity
of the toroidal mean magnetic field is directed downward.
When $\alpha_{12}< 0$, but $|\alpha_{12}| > \gamma$, the situation is
opposite, i.e., the effective pumping velocity of the toroidal mean
magnetic field is directed upward, while the effective pumping velocity
of the poloidal mean magnetic field is directed downward.
Therefore, the effective pumping velocity, $\bm{\gamma}$,
as well as the off-diagonal components of the tensor $\alpha_{ij}$, result in a separation of toroidal and poloidal components of the
mean magnetic field.
This effect is very important for large-scale dynamo action in shear
flow turbulence.

Another reason for the different pumping velocity
of toroidal and poloidal components of the
mean magnetic field is a combination of the effects of rotation
and stratification on small-scale turbulence.
The effect of the separation of toroidal and poloidal
components of the mean magnetic field was early
identified in analytic calculations
of rotating stratified turbulence in \cite{Kit91,KR03},
confirmed in DNS of rotating stratified convection
\cite{Ossen02,Kap06}, and included in numerical
mean-field modeling of the solar dynamo in \cite{KKT06}.
Note also that a nonlinear feedback of the mean magnetic field to
turbulent fluid flow causes a different pumping velocity
of toroidal and poloidal components of the
mean magnetic field \cite{RK04}.
The latter effect was included in numerical mean-field modeling
of the solar dynamo in \cite{ZSR06}.

In summary, we have developed a mean-field theory for a pumping effect
of the mean magnetic field in homogeneous helical turbulence
with imposed large-scale shear.
In our analysis we use the quasi-linear approach, the path-integral
technique and tau-relaxation approximation, which allow us to
determine all components of the $\alpha$ tensor in different ranges of
hydrodynamic and magnetic Reynolds numbers.
The pumping effect depends on the $\alpha$ effect and on shear.
Using DNS and the kinematic test-field method we were able to
determine all components of the $\alpha$ tensor from numerical
simulations
of sheared helical turbulence.
The major part of the
numerical results for the effective pumping velocity,
the diagonal and off-diagonal components of the $\alpha$ tensor
are in a good agreement with the theoretical results.
However, the numerical results for $\alpha_{12}$ suggest
that there is no dependence of the off-diagonal component
on shear in contradiction with
the analytical result.
In addition, according to the numerical results
$\alpha_{12}(\Rey)$ is proportional
to $\Rey$ for small $\Rem$, while the theory yields
$\alpha_{12} \propto \Rey^2$.
On the other hand, the change of the sign of $\alpha_{12}$
from positive for small $\Rem$ to negative for large
$\Rem$ observed in DNS is in
agreement with the theoretical predictions.

\begin{acknowledgements}
Numerous illuminating discussions with Alexander Schekochihin
on the shearing waves approach are kindly acknowledged.
The numerical simulations were performed with the supercomputers
hosted by CSC -- IT Center for Science in Espoo, Finland, who are
administered by the Finnish Ministry of Education. Financial support
from the Academy of Finland grant Nos.\ 136189, 140970,
the Swedish Research Council grant 621-2007-4064, COST Action MP0806, and the European Research Council under the AstroDyn Research Project
227952 are acknowledged.
The authors acknowledge the hospitality of NORDITA.
\end{acknowledgements}

\appendix
\section{The integrals of the Green functions}
\label{Append-A}

For the integration in $\omega$-space in the case of small
magnetic and hydrodynamic Reynolds
numbers we used the following integrals in
\Eqs{new-vel-small-Rm}{new-alpha-small-Rm}:
\begin{eqnarray*}
I_0(k) &=& \int G_\eta \,G_\nu \,G^*_\nu \, d\omega
= {\pi \over \nu \, (\nu + \eta) \, k^4} ,
\\
I_1(k) &=& \int G^2_\eta \,G^2_\nu \,G^*_\nu \, d\omega
= {\pi \over 2 \, \nu^2 \, (\nu + \eta)^2 \, k^8} ,
\\
I_2(k) &=& \int G^2_\eta \,G_\nu \,(G^*_\nu)^2 \, d\omega
= {\pi \, (5 \nu + \eta) \over 2 \, \nu^2 \,
(\nu + \eta)^3 \, k^8} ,
\\
I_3(k) &=& \int G_\eta \,G_\nu \,(G^*_\nu)^3
\, d\omega = {\pi \over 4 \, \nu^3 \, (\nu + \eta)^3\,k^8}
\\
&& \times \, [2 \nu (\nu + \eta)+ (\nu + \eta)^2
+ 4 \nu^2] ,
\\
I_4(k) &=& \int G_\eta \,G_\nu \,(G^*_\nu)^2
\, d\omega = {\pi \, (3 \nu + \eta) \over 2 \,
\nu^2 \, (\nu + \eta)^2 \, k^6} ,
\\
I_5(k) &=& \int G_\eta \,G^2_\nu \,G^*_\nu
\, d\omega = {\pi \over 2 \, \nu^2
\, (\nu + \eta) \, k^6} ,
\\
I_6(k) &=& \int G_\eta \,G^3_\nu \,G^*_\nu
\, d\omega = {\pi \over 4 \, \nu^3
\, (\nu + \eta) \, k^8} ,
\\
I_7(k) &=& \int G^3_\eta \,G_\nu \,G^*_\nu
\, d\omega = {\pi \over \nu \,
(\nu + \eta)^3 \, k^8} ,
\\
I_8(k) &=& \int G^2_\eta \,G_\nu \,G^*_\nu
\, d\omega = {\pi \over \nu \, (\nu + \eta)^2
\, k^6} .
\end{eqnarray*}

\section{Derivation of Eqs.~(\ref{M1}) and~(\ref{M5}) in
path-integral approach}
\label{Append-B}

To derive Eq.~(\ref{M1}) we use an exact solution of the
induction equation with an initial condition $ {\bm B}(t=t_0,{\bm x})
= {\bm B}(t_0,{\bm x}) $ in the form of the Feynman-Kac formula:
\begin{eqnarray}
B_{i}(t,{\bm x})  = \langle G_{ij}(t,t_0,\bec{\xi}) \,
B_{j}(t_0,\bec{\xi}) \rangle_{\bm w} \;,
\label{T5}
\end{eqnarray}
and assume that
\begin{eqnarray}
B_{j}(t_0, \bec{\xi}) = \int \exp(i \bec{\xi} \cdot {\bm q})
B_{j}(t_0, {\bm q}) \,d{\bm q} \; .
\label{CC8}
\end{eqnarray}
Substituting Eq.~(\ref{CC8}) into Eq.~(\ref{T5}) we obtain
\begin{eqnarray}
B_{i}(t, {\bm x}) &=& \int \langle G_{ij}(t,t_0,\bec{\xi})
\, \exp[i \bec{\hat \xi} \cdot {\bm q}] \rangle_{\bm w}
\nonumber\\
& & \times \, B_{j}(t_0, {\bm q}) \, \exp(i {\bm q} \cdot {\bm x})
\,d{\bm q} \;,
\label{C8}
\end{eqnarray}
where $\bec{\hat \xi}=\bec{\xi} - {\bm x}$.
In Eq.~(\ref{C8}) we expand the function $\exp[i \bec{\hat \xi}
\cdot {\bm q}]$ in Taylor series at ${\bm q} = 0$:
\begin{eqnarray*}
\exp(i \bec{\hat \xi} \cdot {\bm q}) = \sum_{k=0}^{\infty}
\, {1 \over k!} \, (i \bec{\hat \xi} \cdot {\bm q})^{k},
\end{eqnarray*}
and use the identity:
\begin{eqnarray*}
\bec{\nabla}^{k} \exp(i {\bm x} \cdot {\bm q}) = (i {\bm q})^{k}
\exp(i {\bm x} \cdot {\bm q})  .
\end{eqnarray*}
This allows us to rewrite Eq.~(\ref{C8}) as follows:
\begin{eqnarray}
B_{i}(t, {\bm x}) &=& \langle G_{ij}(t,t_0,\bec{\xi})
\Big[\sum_{k=0}^{\infty} \, {1 \over k!} \, (\bec{\hat \xi} \cdot
\bec{\nabla})^{k}\Big] \rangle_{\bm w}
\nonumber\\
& & \times \int B_{j}(t_0, {\bm q}) \exp(i {\bm q} \cdot {\bm x})
\,d{\bm q} \; .
\label{BC8}
\end{eqnarray}
After the inverse Fourier transformation, $B_{j}(t_0, {\bm x}) =
\int B_{j}(t_0, {\bm q}) \exp(i {\bm q} \cdot {\bm x}) \,d{\bm q}$,
in Eq.~(\ref{BC8}) we obtain Eq.~(\ref{M1}).
Equation~(\ref{CC8}) can be formally considered
as an inverse Fourier transformation of the function $B_{j}(t_0,
\bec{\xi})$. Equation~(\ref{M1}) has been also derived
by a rigorous method, using the Feynman-Kac formula and
Cameron-Martin-Girsanov theorem (see \cite{KRS02}).

Averaging Eq.~(\ref{M1}) over the random velocity field yields
the equation for the mean magnetic field
\begin{eqnarray}
\meanB_{i}((m + 1) \tau, {\bm x}) &=& \langle \langle
G_{ij}(t,s,\bec{\xi}) \, \exp(\bec{\hat \xi} \cdot \bec{\nabla})
\rangle \rangle_{\bm w}
\nonumber\\
&& \times \meanB_{j}(m \tau, {\bm x}) ,
\label{D6}
\end{eqnarray}
where the angular brackets $ \langle \cdot \rangle $ denote the
ensemble average over the random velocity field. Now we use the identity
\begin{eqnarray}
\meanB_{i}(t+\tau,{\bm x}) = \exp\biggl(\tau {\partial \over
\partial t} \biggr) \meanB_{i}(t,{\bm x}) ,
\label{D8}
\end{eqnarray}
which follows from the Taylor expansion
\begin{eqnarray*}
f(t + \tau) = \sum_{m=1}^{\infty} \biggl(\tau {\partial \over
\partial t} \biggr)^{m} f(t) = \exp \biggl(\tau {\partial \over
\partial t} \biggr) {f(t) \over m!}  .
\end{eqnarray*}
Therefore, Eqs. (\ref{D6})--(\ref{D8}) yield
\begin{eqnarray}
&& \exp\biggl(\tau {\partial \over \partial t} \biggr)
\meanB_{i}(t,{\bm x}) =  (\overline{G}_{ij} + \overline{G}_{ij} \overline{\xi}_{m}
\nabla_m + A_{ijm} \nabla_m
\nonumber\\
&& \quad + C_{ijmn} \nabla_m \nabla_n) \meanB_{j} \equiv
\exp(\tau \hat L) \, \meanBB ,
\label{D9}
\end{eqnarray}
where $\overline{G}_{ij} = \langle\langle G_{ij} \rangle \rangle_{\bm w} =
\delta_{ij} + \meanU_{i,j} \, \tau + O[(S\tau)^2] ,$
$\, \overline{\xi}_{i} = \langle\langle \hat \xi_{i} \rangle \rangle_{\bm w}
= - \meanU_{i} \, \tau + O[(S\tau)^2] ,$ $\, A_{ijm} =
\langle\langle \hat \xi_m G_{ij} \rangle \rangle_{\bm w} ,$
$ \, C_{ijmn} = \langle\langle \hat \xi_m
\hat \xi_n G_{ij} \rangle \rangle_{\bm w} ,$ and
we introduced the operator $\hat L ,$ which allows us to reduce the
integral equation~(\ref{D6}) to a partial differential equation.
Indeed, Eq.~(\ref{D9}), which is rewritten in the form
\begin{eqnarray}
\exp \biggl[\tau \biggl(\hat L - {\partial \over \partial t}
\biggr) \biggr] \meanBB = \meanBB ,
\label{D11}
\end{eqnarray}
reduces to
\begin{eqnarray}
{\partial \meanBB \over \partial t} = \hat L \meanBB  .
\label{D12}
\end{eqnarray}
Taylor expansion of the function $\exp(\tau \hat L)$ reads
\begin{eqnarray}
\exp(\tau \hat L) = \hat E + \tau \hat L + (\tau \hat L)^2 / 2 +
... ,
\label{D14}
\end{eqnarray}
where $\hat E $ is the unit operator. Thus, Eqs.~(\ref{D9})
and~(\ref{D14}) yield
\begin{eqnarray}
\hat L &\equiv& L_{ij} = {1 \over \tau} (\overline{G}_{ij} - \delta_{ij}
+  \overline{\xi}_{m} \overline{G}_{ij} \nabla_m + A_{ijm} \nabla_m)
\nonumber\\
&&+ D_{ijmn} \nabla_m \nabla_n + O(\nabla^3) ,
\label{D15}
\end{eqnarray}
where $ D_{ijmn} = (C_{ijmn} - A_{ikm} A_{kjn}) / 2 \tau$.
This yields Eq. (\ref{M5}).

\section{Orr-Kelvin random shearing waves for small hydrodynamic Reynolds numbers}
\label{Append-C}

We explain here the details that led to the derivation of \Eqs{M11}{M12}.
We seek the solutions of the linearized Eq.~(\ref{A1}) for incompressible velocity
field ${\bm u}$ as superpositions of the Orr-Kelvin shearing waves:
\begin{eqnarray}
{\bm u}(t,{\bm r}) = \int {\bm u}(t,{\bm k}_0) \exp [i {\bm k}(t)
\cdot {\bm r}] \, d{\bm k}_0,
\label{L100}
\end{eqnarray}
(see, e.g., \cite{K1887,Oa1907,Ob1907,SKR08}), where
${\bm k}_0=(k_{x0}, k_y, k_z)$, $\, {\bm k}(t)=(k_{x0}-S k_y t,k_y,k_z)$
and we neglected weak Lorentz force.
The amplitudes of the shearing waves satisfy the following equations:
\begin{eqnarray}
{\partial u_x(t,{\bm k}_0) \over \partial t} &=& \left[2 S {k_y k_x(t)
\over k^2(t)} -\nu k^2(t) \right] \, u_x(t,{\bm k}_0) + f_x,
\nonumber\\
\label{L2}\\
{\partial u_z(t,{\bm k}_0) \over \partial t} &=& 2 S {k_y k_z \over
k^2(t)} \, u_x(t,{\bm k}_0)- \nu k^2(t) \, u_z(t,{\bm k}_0) + f_z .
\nonumber\\
\label{L3}
\end{eqnarray}
These equations were obtained by taking twice {\bf curl} of Eq.~(\ref{A1}). Equations~(\ref{L2}) and~(\ref{L3})
have explicit solutions:
\begin{eqnarray}
u_x(t,{\bm k}_0) &=& {1 \over k^2(t)} \, \int_0^t dt' \,
\tilde G_\nu(t,t') \, k^2(t') \, f_x(t',{\bm k}_0),
\nonumber\\
\label{L4}\\
u_z(t,{\bm k}_0) &=& u_z^{(1)}(t,{\bm k}_0) + u_z^{(2)}(t,{\bm k}_0),
\label{L5}\\
u_y(t,{\bm k}_0) &=& - {1 \over k_y} \, \left[k_x(t) \, u_x(t,{\bm k}_0)
+ k_z \, u_z(t,{\bm k}_0) \right],
\label{L11}\\
u_z^{(1)}(t,{\bm k}_0) &=& \int_0^t dt' \, \tilde G_\nu(t,t')
\, f_z(t',{\bm k}_0) ,
\label{L6}\\
u_z^{(2)}(t,{\bm k}_0) &=& 2S \, k_y k_z \, \int_0^t dt' \,
{\tilde G_\nu(t,t') \over k^2(t')} \, u_x(t',{\bm k}_0),
\label{L7}
\end{eqnarray}
where $\tilde G_\nu(t,t') = \exp\left[-\nu\int_{t'}^t \,dt'' k^2(t'')\right]$.
Equations~(\ref{L4})--(\ref{L7}) for a white-in-time forcing yield
the following formulas for non-instantaneous two-point correlation functions:
\begin{eqnarray}
&& \langle u_{x}(t,{\bm k}_{0}) \, u_{z}^{*(1)}(t',{\bm k}_{0}) \rangle = \tilde G_\nu(t,t') \, {k^{2}(t') \over k^{2}(t)}
\nonumber\\
&& \quad \quad\quad \quad\quad \times \, \, \langle u_{x}(t',{\bm k}_0) \, u_{z}^{*(1)}(t',{\bm k}_0) \rangle ,
\label{L9}\\
&& \langle u_{z}^{(1)}(t,{\bm k}_{0}) \, u_{x}^*(t',{\bm k}_{0}) \rangle = \tilde G_\nu(t,t')
\nonumber\\
&& \quad \quad\quad \quad\quad \times \,  \, \langle u_{z}^{(1)}(t',{\bm k}_0) \, u_{x}^*(t',{\bm k}_0) \rangle ,
\label{L10}\\
&& \langle u_{x}(t,{\bm k}_{0}) \, u_{z}^{*(2)}(t',{\bm k}_{0})
\rangle = 2S \, k_{y} k_{z} \, \int_0^{t'} dt''  \, {\tilde G_\nu(t',t'') \over k^{2}(t'')}
\nonumber\\
&& \quad \quad\quad \quad\quad \times  \, \langle u_{x}(t,{\bm k}_0)
\, u_{x}^*(t'',{\bm k}_0) \rangle ,
\label{L12}\\
&& \langle u_{z}^{(2)}(t,{\bm k}_{0}) \, u_{x}^*(t',{\bm k}_{0})
\rangle = 2S \, k_{y} k_{z} \, \int_0^{t} dt''  \, {\tilde G_\nu(t,t'') \over k^{2}(t'')}
\nonumber\\
&& \quad \quad\quad \quad\quad \times  \, \langle u_{x}(t'',{\bm k}_0)
\, u_{x}^*(t',{\bm k}_0) \rangle ,
\label{L13}
\end{eqnarray}
where for $t'' < t'$
\begin{eqnarray}
&& \langle u_{x}(t'',{\bm k}_{0}) \, u_{x}^{*}(t',{\bm k}_{0}) \rangle = \tilde G_\nu(t',t'') \, {k^{2}(t'') \over k^{2}(t')}
\nonumber\\
&& \quad \quad\quad \quad \quad \times \,  \, \langle u_{x}(t'',{\bm k}_0) \, u_{x}^{*}(t'',{\bm k}_0) \rangle ,
\label{L8}
\end{eqnarray}
and for $t'' > t'$
\begin{eqnarray}
&& \langle u_{x}(t'',{\bm k}_{0}) \, u_{x}^{*}(t',{\bm k}_{0}) \rangle = \tilde G_\nu(t'',t') \, {k^{2}(t') \over k^{2}(t'')}
\nonumber\\
&& \quad \quad\quad \quad \quad \times \,  \, \langle u_{x}(t',{\bm k}_0) \, u_{x}^{*}(t',{\bm k}_0) \rangle .
\label{L14}
\end{eqnarray}


\begin{thebibliography}{99}

\bibitem {M78} H. K. Moffatt, {\em Magnetic Field Generation in
Electrically Conducting  Fluids} (Cambridge University Press, New
York, 1978).

\bibitem {P79} E. Parker, {\it Cosmical Magnetic Fields} (Oxford
University Press,  New York,  1979).

\bibitem {KR80} F. Krause, and K. H. R\"{a}dler, {\it Mean-Field
Magnetohydrodynamics and  Dynamo Theory} (Pergamon, Oxford, 1980).

\bibitem {ZRS83} Ya. B. Zeldovich, A. A. Ruzmaikin, and D. D.
Sokoloff, {\em Magnetic Fields in Astrophysics} (Gordon and Breach,
New York, 1983).

\bibitem {RSS88} A. Ruzmaikin, A. M. Shukurov, and D. D.
Sokoloff, {\it Magnetic Fields of Galaxies} (Kluwer Academic,
Dordrecht, 1988).

\bibitem {O03}
M. Ossendrijver,  Astron. Astrophys. Rev. {\bf 11}, 287 (2003).

\bibitem {BS05}
A. Brandenburg and K. Subramanian, Phys. Rept. {\bf 417}, 1 (2005).

\bibitem {RK03}
I. Rogachevskii and N. Kleeorin, Phys. Rev. E {\bf 68}, 036301 (2003).

\bibitem {RK04}
I. Rogachevskii and N. Kleeorin, Phys. Rev. E {\bf 70}, 046310 (2004).

\bibitem {EKR03}
T. Elperin, N. Kleeorin and I. Rogachevskii,
Phys. Rev. E {\bf 68}, 016311 (2003).

\bibitem {EGKR07}
T. Elperin, I. Golubev, N. Kleeorin and I. Rogachevskii, Phys. Rev. E {\bf 76}, 066310 (2007).

\bibitem {BR05}
A. Brandenburg, Astrophys. J. {\bf 625},
539-547 (2005).

\bibitem {YHS08}
T. A. Yousef, T. Heinemann, A. A. Schekochihin, N. Kleeorin, I. Rogachevskii, A. B. Iskakov, S. C. Cowley, J. C. McWilliams, Phys. Rev. Lett. {\bf 100}, 184501 (2008).

\bibitem {YHR08}
T. A. Yousef, T. Heinemann, F. Rincon, A. A. Schekochihin, N. Kleeorin, I. Rogachevskii, S. C. Cowley, J. C. McWilliams, Astron. Nachr. {\bf 329}, 737 (2008).

\bibitem {BRRK08}
A. Brandenburg, K.-H. R\"adler, M. Rheinhardt, P. J. K\"apyl\"a,
Astrophys. J. {\bf 676}, 740 (2008).

\bibitem {KKB08}
P. J. K\"apyl\"a, M. J. Korpi and A. Brandenburg, Astron. Astrophys.
{\bf 491}, 353 (2008).

\bibitem {HP09}
D. W. Hughes and M. R. E. Proctor, Phys. Rev. Lett. {\bf 102}, 044501 (2009).

\bibitem {RS06}
K.-H. R\"{a}dler and R. Stepanov, Phys. Rev. E {\bf 73}, 056311
(2006).

\bibitem {RK06}
G. R\"{u}diger and L. L. Kitchatinov, Astron. Nachr. {\bf 327}, 298
(2006).

\bibitem {KR08}
N. Kleeorin and I. Rogachevskii, Phys. Rev. E {\bf 77}, 036307 (2008).

\bibitem {SS09}
S. Sridhar and K. Subramanian, Phys. Rev. E {\bf 79}, 045305(R) (2009).

\bibitem {SS10}
S. Sridhar and N. K. Singh,  J. Fluid Mech. {\bf 664}, 265 (2010).

\bibitem {SKR08} T. Heinemann, A. A. Schekochihin,
and J. C. McWilliams, E-print: arXiv:0810.2225.

\bibitem {KMB09}
P. J. K\"{a}pyl\"{a}, D. Mitra, A. Brandenburg, Phys. Rev. E {\bf 79}, 016302 (2009).

\bibitem {MKB09}
D. Mitra,  P. J. K\"{a}pyl\"{a}, R. Tavakol,
A. Brandenburg, Astron. Astroph.  {\bf 495}, 1 (2009).

\bibitem {Kit91}
L. L. Kitchatinov, Astron. Astrophys. {\bf 243}, 483 (1991).

\bibitem {RKR03}
K.-H. R\"{a}dler, N. Kleeorin and I. Rogachevskii, Geophys.
Astrophys. Fluid Dynamics {\bf 97}, 249 (2003).

\bibitem {RS75} P. H. Roberts and A. M. Soward,  Astron. Nachr. {\bf
296}, 49 (1975).

\bibitem {KR82} N. Kleeorin, and A. Ruzmaikin,
Magnetohydrodynamics {\bf No. 2}, 17 (1982).

\bibitem {GD94} A. V. Gruzinov, and P. H. Diamond, Phys.
Rev. Lett., {\bf 72}, 1651 (1994).

\bibitem {GD95} A. V. Gruzinov, and P. H. Diamond, Phys. Plasmas {\bf 2}, 1941
(1995).

\bibitem {KR99}
N. Kleeorin and I. Rogachevskii, Phys. Rev. E {\bf 59}, 6724 (1999).

\bibitem {O70}  S. A. Orszag, J. Fluid Mech. {\bf 41}, 363 (1970).

\bibitem {MY75} A. S. Monin and A. M. Yaglom, {\it Statistical Fluid
Mechanics}  (MIT Press, Cambridge, Massachusetts, 1975), Vol. 2.

\bibitem {Mc90} W. D. McComb, {\it The Physics of Fluid Turbulence}
(Clarendon,  Oxford, 1990).

\bibitem {PFL76} A. Pouquet, U. Frisch, and J. Leorat, J. Fluid Mech.
{\bf 77}, 321 (1976).

\bibitem {KRR90}
N. Kleeorin, I. Rogachevskii, and A. Ruzmaikin, Zh. Eksp. Teor. Fiz.
{\bf 97}, 1555 (1990) [Sov. Phys. JETP {\bf 70}, 878 (1990)].

\bibitem {BF02} E. G. Blackman and G. B. Field,  Phys. Rev. Lett.
{\bf 89}, 265007 (2002).

\bibitem {BF03} E. G. Blackman and G. B. Field,
Phys. Fluids {\bf 15}, L73 (2003).

\bibitem {FB02} G. B. Field and E. G. Blackman,  Astrophys. J.
{\bf 572}, 685 (2002).

\bibitem {BK04}
A. Brandenburg, P. K\"{a}pyl\"{a}, and A. Mohammed, Phys. Fluids
{\bf 16}, 1020 (2004).

\bibitem {BSM05}
A. Brandenburg and K. Subramanian, Astron. Astrophys. {\bf 439}, 835
(2005).

\bibitem {BS07}
A. Brandenburg and K. Subramanian, Astron. Nachr. {\bf 328}, 507
(2007).

\bibitem {SSB07}
S. Sur, K. Subramanian and A. Brandenburg, Monthly Notices Roy.
Astron. Soc. {\bf 376}, 1238 (2007).

\bibitem {RK07}
I. Rogachevskii and N. Kleeorin, Phys. Rev. E {\bf 76}, 056307 (2007).

\bibitem {DM84}
P. Dittrich, S. A. Molchanov, A. A. Ruzmaikin and D. D.
Sokoloff, Astron. Nachr. {\bf 305}, 119 (1984).

\bibitem {KRS02} N. Kleeorin, I. Rogachevskii and D. Sokoloff,
Phys. Rev. E {\bf 65}, 036303 (2002).

\bibitem {K1887}
Kelvin (W. Thomson), Phil. Mag. {\bf 24} (5), 188 (1887).

\bibitem {Oa1907}
W. M. Orr, Proc. R. Irish Acad. A {\bf 27}, 9 (1907).

\bibitem {Ob1907}
W. M. Orr, Proc. R. Irish Acad. A {\bf 27}, 69 (1907).

\bibitem {RK97}
I. Rogachevskii and N. Kleeorin, Phys. Rev. E {\bf 56}, 417 (1997).

\bibitem {B01} A. Brandenburg,  Astrophys. J.
{\bf 550}, 625 (2001).

\bibitem {HBD04} N. E. L. Haugen, A. Brandenburg, and W. Dobler,
Phys. Rev. E {\bf 70}, 016308 (2004).

\bibitem {Sch05}
M. Schrinner, K.-H. R\"adler, D. Schmitt, M. Rheinhardt, and
U. Christensen,
Astron. Nachr. {\bf 326}, 245 (2005).

\bibitem {Sch07}
M. Schrinner, K.-H. R\"adler, D. Schmitt,  M. Rheinhardt, and
U. R. Christensen,
Geophys. Astrophys. Fluid Dyn. {\bf 101}, 81 (2007).

\bibitem {KR03}
N. Kleeorin and I. Rogachevskii, Phys. Rev. E {\bf 67}, 026321 (2003).

\bibitem {Ossen02}
M. Ossendrijver, M. Stix, A. Brandenburg and G. R\"udiger, Astron. Astrophys. {\bf 394}, 735 (2002).

\bibitem {Kap06}
P. J. K\"{a}pyl\"{a}, M. J. Korpi, M. Ossendrijver and M. Stix,
Astron. Astrophys. {\bf 455}, 401 (2006).

\bibitem {KKT06}
P. J. K\"{a}pyl\"{a}, M. J. Korpi and I. Tuominen, Astron. Nachr.
{\bf 327}, 884 (2006).

\bibitem {ZSR06}
H. Zhang, D. Sokoloff, I. Rogachevskii, D. Moss, V. Lamburt,
K. Kuzanyan and N. Kleeorin, Mon. Not. R. Astron. Soc. {\bf 365}, 276 (2006).

\end{thebibliography}
\end{document}